# Kinetic Electron Cooling in Magnetic Nozzles: Experiments and Modeling

June Young Kim[1*], Kyoung-Jae Chung[1], Kazunori Takahashi[2], Mario Merino[3], and Eduardo Ahedo[3]

[1]*Department of Nuclear Engineering, Seoul National University, Seoul, Korea*

[2]*Department of Electrical Engineering, Tohoku University, Japan*

[3]*Equipo de Propulsión Espacial y Plasmas (EP2), Universidad Carlos III de Madrid, Madrid, Spain*

## ABSTRACT

As long-distance space travel requires propulsion systems with greater operational flexibility and lifetimes, there is a growing interest in electrodeless plasma thrusters that offer the opportunity of improved scalability, larger throttleability, running on different propellants, and limit device erosion. The majority of electrodeless designs rely on a magnetic nozzle (MN) for the acceleration of the plasma, which has the advantage of utilizing the expanding electrons to neutralize the ion beam without the additional installation of a cathode. The plasma expansion in the MN is nearly collisionless, and a fluid description of electrons requires a non-trivial closure relation. Kinetic electron effects, and in particular electron cooling, play a crucial role in various physical phenomena such as energy balance, ion acceleration, and particle detachment. Based on the experimental and theoretical studies conducted in recognition of this importance, the fundamental physics of the electron cooling mechanism revealed in MNs and magnetically expanding plasma are reviewed. Especially, recent approaches from the kinetic point of view are discussed, and our perspective on the future challenges of electron cooling and the relevant physical subject of MN is presented.

[*]Corresponding author: E-mail: ptcbcg@gmail.com, beacoolguy@snu.ac.kr

**TABLE OF CONTENTS**

# I. INTRODUCTION

In the development of technology for deep-space exploration of long-duration space missions, space propulsion requires higher thrust efficiency and longer life-time. Magnetic nozzle (MN)-based devices are attracting attention as next-generation electric thrusters with advantages, such as contactless and electrodeless plasma acceleration, avoiding the erosion of the device, enabling a higher throttleablity range, and facilitating the use of alternative propellants (Ahedo, 2011a; Merino and Ahedo, 2017; Sutton, 2017; Arfeive and Breizman, 2005; Levchenko *et al*., 2020; Takahashi, 2020; Sheppard and Little, 2020). 3D steerable MNs have also been proposed and demonstrated for the simplified adjustment of ion-beam trajectory and thrust vector control (Merino and Ahedo, 2018; Merino and Ahedo, 2017; Charles *et al*., 2008; Imai and Takahashi, 2021; Takahashi and Imai, 2022). The MN has been recognized as the acceleration stage in the development of next-generation space plasma thrusters such as the applied-field magneto plasma dynamic thruster (AF-MPD) (Choueiri, 1998; Andrenucci, 2010; Kodys and Choueiri, 2005), the electrodeless propulsions of helicon plasma thruster (HPT) (Ziemba *et al*., 2005; Takahashi, 2019; Takahashi *et al*., 2011), electron cyclotron resonance plasma thruster (ECRT) (Sercel, 1987; Correyero *et al*., 2019; Vialis *et al*., 2018), and variable specific impulse plasma rocket (VASIMIR) (Chang-Diaz, 2000 a; 2000b; Arefiev and Breizman, 2004) using alternative current ranging from radio-frequency (RF) to microwave (MW) power source. The proposed electric thrusters have different characteristics from the plasma generation and heating viewpoint, but the physics of the quasineutral, quasi-collisionless plasma expansion in the MN are essentially common for all of them: the diverging magnetic field confines the plasma radially and helps convert perpendicular energy into parallel energy, while the thermal energy available in the electrons is converted to ion kinetic energy via the self-consistent electrostatic field. The electron response, and in particular their temperature, plays a fundamental role in the set-up of the electrostatic field in the plume, which is responsible for ion acceleration, modification of the magnetic field structure, and plasma detachment (Deline et al., 2009; Ahedo and Merino, 2011; Ahedo and Merino, 2012; Merino and Ahedo, 2014; Hooper, 1993; Breizman *et al*., 2008; Olsen et al., 2014; Ebersohn *et al*., 2012; Deline *et al*., 2009; Little and Choueiri, 2019; Little and Choueiri, 2013; Takahashi and Ando, 2017b). Accordingly, for the development of MN-based devices, it is essential to understand the kinetics of electron cooling along the divergent magnetic field.

A collisionless, magnetically expanding plasma has quite complex physical elements, and overlooking the kinetics of electrons (e.g., by using a single fluid approach with either isothermal or polytropic closures) can dictate the wrong directions in device development (Kaganovich *et al*., 2020). Theoretically, the invariants of motions in electric and magnetic fields (the conservation of the energy and the magnetic moment) result in a complex electron velocity distribution function (EVDF) in the MNs, e.g., an anisotropic and partially depleted (Martinez-Sanchez *et al*., 2015; Sanchez-Arriaga *et al*., 2018; Merino *et al*., 2018; Ahedo *et al*., 2020; Merino *et al*., 2021). Electrons are classified into free, reflected, and doubly-trapped populations according to the effective potential that defines their motion (Martinez-Sanchez *et al*., 2015) The doubly-trapped electron population, whose trajectories are disconnected from the plasma source, depends on the transient plume setup process and the weak collisionality that may exist in the plasma. In accordance with the complexity, the recent modeling results provide clues on the interpretation of the thermodynamic state of electrons, which is far from local equilibrium, emphasizing that the heat-flux of anisotropic energy distribution has a dominant role in the electron energy equation (Merino *et al*., 2018; Ahedo *et al*., 2020; Merino *et al*., 2021; Hu *et al*., 2021).

Measurements of the electron velocity distribution function (EVDF) have revealed kinetic behavior of the electrons in the MN. The EVDF measured in the source has shown the depleted tail at the break energy corresponding to the potential drop, i.e., the EVDF has the high-temperature, low-energy population and the low-temperature, high-energy population. The former is trapped by the electric field, while the latter can overcome the electric field and neutralize the supersonic ion beam (Takahashi *et al*., 2011; Plihon *et al*., 2007; Takahashi and Ando, 2017). Since the energization of the electrons is due to the RF heating near the antenna, the spatial mapping of the EVDFs has also clarified some kinetic aspects of the electron transport dynamics and structural formation, e.g., in Refs. (Takahashi *et al*., 2009; Charles, 2010; Takahashi *et al*., 2017a; Gulbrandsen and Fredriksen, 2017; Ghosh *et al*., 2018). Recent experimental studies have tried to determine the thermodynamic state of electrons in electric and magnetic fields (Kim *et al*., 2018; Kim *et al*., 2019; Kim *et al*., 2021a; 2021b; Little and Choueiri, 2016; Takahashi *et al*., 2018; Takahashi *et al*., 2020; Sheehan *et al*., 2014, Zhang *et al*., 2016a; Lafleur *et al*., 2015; Boni *et al*., 2022; Vinci *et al*., 2022). Although more effort is required to experimentally prove the anisotropic behavior of electrons predicted by the theoretical works, the analysis of the spatial distribution of electron properties gives rise to a major contribution to the establishment and verification of the theory for the electron cooling process. Recent studies (Kim *et al*., 2018; Kim *et al*., 2019; Kim *et al*., 2021a; 2021b; Takahashi *et al*., 2018; Takahashi *et al*., 2020) have succeeded in finding out that each electron group can have a different thermodynamic state based on the classification of electrons suggested by the models, and this advancement in knowledge has presented a new perspective on the anomalous quasi-isothermal behavior of electrons in the divergent magnetic field observed in the universe and laboratory plasmas as well as improving the performance of MN devices. In the development of electrodeless propulsion, the consensus found between theoretical and experimental studies demands a summary of the essential topic of electron cooling that has been explored for about 50 years (Litvinov, 1971; Andersen *et al*., 1969, Raadu, 1979; Kuriki and Okada, 1970; Arefiev and Breizman, 2008). We believe that this work will be a stepping-stone in the expansion of the research field to various topics scattered for improving the performance of MNs and for physical understanding. In this regard, we provide a review of the kinetic features of electron cooling the fundamental principle of the MN.

The rest of the paper is structured as follows. Section II presents and discusses relevant experimental results to understand electron cooling and kinetic effects in a MN. Section III summarizes the basic fluid model of the plasma expansion in a MN, and then examines electron cooling from a theoretical viewpoint, reviewing recent modeling and numerical results. Finally, Section IV gathers the main conclusions and outlines the open challenges on this matter.

## II. EXPERIMENTAL APPROACH TO ELECTRON THERMODYNAMICS IN MAGNETIC NOZZLE

Experimental environments to elucidate the thermodynamic state of electrons in the MN require the low collisionality, minimized plasma-solid boundary effect, and closed path of magnetic field lines. Based on these requirements, the experimental study on electron thermodynamics is to magnetically expand plasma generated in the source region into a diffusion region having a larger volumetric dimension than the source and to analyze the behavior of the plasma using a (local) polytropic exponent. In recent years, intensive studies on the subject of electron thermodynamics have been carried out in the laboratory [Table 1 and 2]. They have engineering and physical significance in that they present a new perspective on the thermodynamic state of electrons relevant to not only the operating mechanism of magnetic nozzles but also the fundamental physics of space plasmas (Kim *et al*., 2018; Kim *et al*., 2019; Kim *et al*., 2021a; 2021b; Little and Choueiri, 2016; Takahashi *et al*., 2018; Takahashi *et al*., 2020; Sheehan *et al*., 2014; Zhang *et al*., 2016; Lafleur *et al*., 2015; Boni *et al*., 2022; Vinci *et al*., 2022).

On the basis of the diagnostics technique and plasma source technology, the electron cooling rate was investigated in relation to a simple description of the ion acceleration in the MN (Sheehan *et al*., 2014; Zhang *et al*., 2016; Lafleur *et al*., 2015). Then, recent comprehensive experiments have taken into account detailed elements such as the trapped motion, the cross-field diffusion, and the degree of freedom of electrons (Kim *et al*., 2018; Kim *et al*., 2019; Takahashi *et al*., 2018; Takahashi *et al*., 2020; Kim *et al*., 2021a; 2021b). Accordingly, this section emphasizes the sequential flow of experimental research and classifies studies into (1) initial studies that do not consider all the factors (the trapped motion, the cross-field diffusion, and degree of freedom), (2) studies that consider the effect of trapped electrons on thermodynamics, and (3) studies that control the thermodynamic state of electrons by modifying the number of degrees of freedom.

**Table.1** Details of experimental information of representative studies of the electron thermodynamics in magnetic nozzle.

| Reference | Neutral pressure | Source type | Magnetic field | Vacuum chamber or expansion chamber |
|---|---|---|---|---|
| (Sheehan et al., 2014) | 0.1 mTorr | Helicon (6.78 MHz) | Electromagnet (2000 G at nozzle throat) | Vacuum chamber (4.2 m in diameter and 10 m in length) |
| (Lafleur et al., 2015) | 3.8 to 7.5 $\mu$Torr | ECR (2.45 GHz) | Electromagnet (<1000 G inside the source) | Vacuum chamber (1 m in diameter, 4 m in length) |
| (Little and Choueiri, 2016) | 0.02 mTorr | ICP (13.56 MHz) | Electromagnet (peak magnetic field, 150−420 G) | Vacuum chamber (2.4 m in diameter, 7.6 m in length) |
| (Zhang et al., 2016) | 0.3 mTorr | Helicon (13.56 MHz) | Electromagnet (peak magnetic field, 150 G) | Expansion chamber (0.32 m in diameter, 0.3 m in length) |
| (Takahashi et al., 2018) | 0.5 mTorr | DC | Electromagnet (peak magnetic field, 220 G) | Expansion chamber (0.15 m in diameter, 0.5 m in length) |
| (Kim et al., 2018) | 0.45 mTorr | ECR (2.45 GHz) | Electromagnet (450 G at nozzle throat) | Expansion chamber (0.6 m in diameter, 0.66 m in length) |
| (Kim et al., 2019) | 0.4 mTorr | ICP (13.56 MHz) | Electromagnet (70 G at nozzle throat) | Expansion chamber (0.6 m in diameter, 0.66 m in length) |
| (Correyero et al., 2019) | 2.1 to 2.8 $\mu$Torr | ECR (2.45 GHz) | Electromagnet or permanent magnet (fixed at 900 G for both types at the thrust back plate) | Vacuum chamber (0.8 m in diameter, 2 m in length) |
| (Takahashi et al., 2019) | 0.5 mTorr | DC | Electromagnet (peak magnetic field, 264 G) | Expansion chamber (0.15 m in diameter, 0.5 m in length) |
| (Kim et al., 2021a) | 0.4 mTorr | DC | Electromagnet (230 G at nozzle throat) | Expansion chamber (0.6 m in diameter, 0.66 m in length) |
| (Kim et al., 2021b) | 0.4 mTorr | DC | Electromagnet (230 G at nozzle throat) | Expansion chamber (0.6 m in diameter, 0.66 m in length) |
| (Vinci et al., 2022) | 0.7 mTorr | Helicon (13.56 MHz) | Electromagnet (peak magnetic field, 86 ± 3 G) | Expansion chamber (0.3 m in diameter, 0.5 m in length) |

Table 2. Details of diagnostics and measured polytropic index

| Reference | Probe tip | Electron temperature | Electron density | Polytropic index | Features |
|---|---|---|---|---|---|
| (Sheehan et al., 2014) | Planar | Simi-log plot of the electron current | Electron saturation current | 1.75 | Set degree of freedom of 2 |
| (Lafleur et al., 2015) | Cylindrical | Druyvesteyn method | – | 1.2−1.55 | No distinctive dependence on flow rate |
| (Little and Choueiri, 2016) | Cylindrical | Simi-log plot of the electron current | Ion saturation current | 1.15 | Near isothermal process |
| (Zhang et al., 2016) | Cylindrical | Druyvesteyn method | | 1.17 | Adiabatic process (non-local thermodynamic equilibrium) |
| (Takahashi et al., 2018) | Cylindrical | Druyvesteyn method | | 1−5/3 | Removed axial electric field |
| (Kim et al., 2018) | Cylindrical | Druyvesteyn method | | 1−5/3 | Spatial variation of polytropic index |
| (Kim et al., 2019) | Cylindrical | Druyvesteyn method | | 1−5/3 | Spatio-temporal variation of polytropic index |
| (Correyero et al., 2019) | Cylindrical | Druyvesteyn method | | 1.23 | Spatial variation of polytropic index |
| (Takahashi et al., 2019) | Cylindrical | Druyvesteyn method | | 1−5/3 | Correlation of cross-field diffusion and polytropic index |
| (Kim et al., 2021a) | Cylindrical | Druyvesteyn method | | 2 | Changes in the degree of freedom by a radial electric field |
| (Kim et al., 2021b) | Cylindrical | Druyvesteyn method | | 1.88 | Verification of reversible process |
| (Vinci et al., 2022) | Cylindrical | Simi-log plot of the electron current or Druyvesteyn method | Ion saturation current | 1.35−1.85 | 2-dimensional measurement of polytropic index |

## A. Basic research on electron thermodynamics

Early studies excluded the in-depth discussion of the thermodynamics of electrons, but rather introduced a polytropic index to provide a simple description of electron cooling in MN devices (Sheehan *et al*., 2014). The experimental study of electron thermodynamics in MNs was revisited during the development of VASIMR. The experiments in the high-vacuum chamber of Ad Astra Company (4.23 m in diameter and 10 m long with a base pressure of $10^{-9}$ Torr) minimized the blocking of the streamline of the magnetic field by the vacuum wall, and thus an experiment in more realistic boundary condition similar to space environment was performed with the helicon source-based MN (VX-200), a prototype electrodeless plasma propulsion device for spacecraft. The main objective of the study was to elucidate the physical meaning of the electron cooling rate and the correlation of plasma potential, and electron temperature and density varying along the divergent magnetic field. In the same context, an essential question was presented whether a current-free double layer observed in some laboratory experiments can be created in a space-like environment.

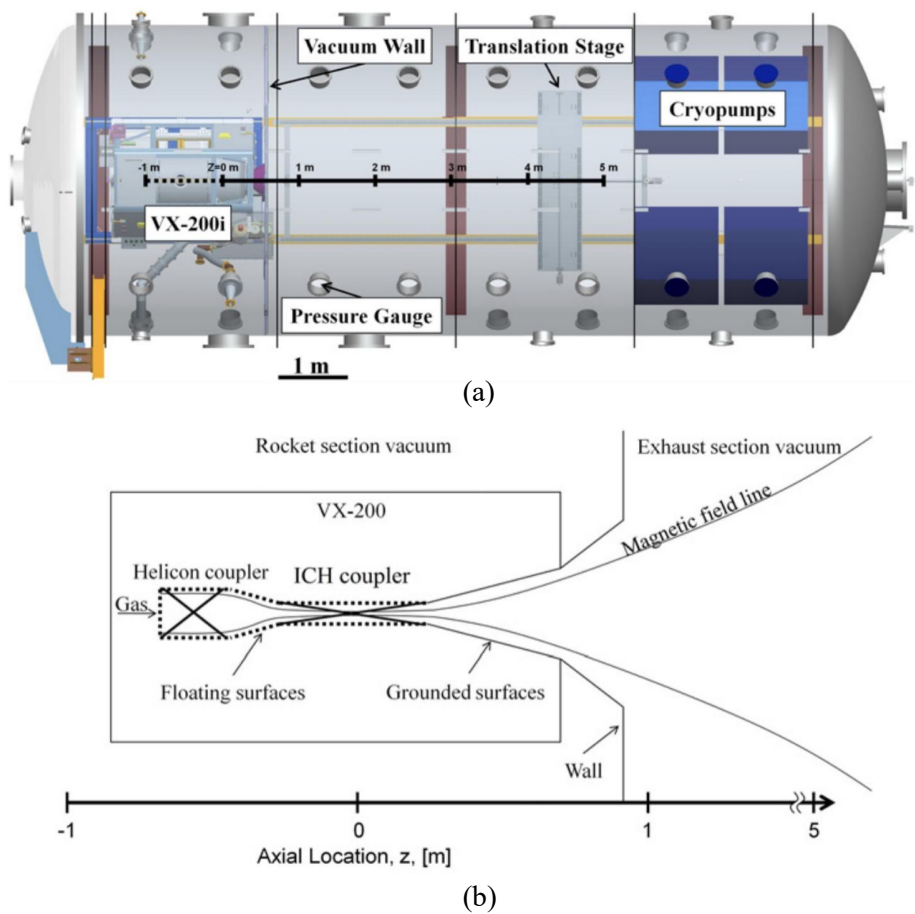

FIG. 1. (a) Schematic illustration of the Ad Astra Rocket Company vacuum chamber (overhead view) (Longmier *et al*., 2011) and (b) diagram of the VX-200 device (Sheehan *et al*., 2014). The ICH coupler

shown in (b) was not used in the experiment of Ref. (Sheehan *et al*., 2014). The operating pressure was almost similar to that of the previous experiments of helicon magnetic nozzle (in the ranges of $10^{-4}$ Torr) while a distinctive difference was the size of the vacuum chamber. Reproduced with permission from Ref. (Longmier *et al*., 2011). Copyright 2011 IOP Publishing. Reproduced with permission from Ref. (Sheehan *et al*., 2014). Copyright 2014 IOP Publishing.

Sheehan *et al*. (Sheehan *et al*., 2014) proposed the correlation of the electron cooling and ambipolar ion acceleration in a MN. They concluded that the plasma system of the MN is adiabatic (i.e., does not exchange energy with its surroundings) in the expansion region so any energy lost by the electrons must be transferred to the ions via the electric field.

They classified three possible theories of electron cooling and relevant ambipolar acceleration mechanism based on previous studies: (1) current-free double layer: a potential gradient equivalent to 10s of electron temperature is generated within a few Debye lengths from the plasma source, and electrons and ion energetic beams are created on the high and low potential side of the double layer, respectively, (2) rarefaction wave theory: a rarefaction wave creates a large potential barrier in the far-downstream region, and electrons lose their energy and become trapped downstream with decreased energy, and (3) adiabatic theory combined with electron momentum equation: generation of an electric field that induces ion acceleration due to electron cooling (adiabatic) process.

Experimental evidence corresponding to the double layer and rarefaction wave theory (such as a strong electric potential layer in a region of tens of Debye lengths near the nozzle throat or at the far-field region) was not observed. Rather, only the relation of the electron temperature $T_e$ and plasma potential $e\phi$ enables discussion of the electron thermodynamics by the relation, $\partial(e\phi)/\partial s = \gamma/(\gamma - 1)\partial T_e/\partial s$, where $s$ is the field-aligned pointing vector and $\gamma$ is the polytropic index of electrons. They considered that $T_e$ measured with the planar probe only collects the temperature component parallel to the magnetic field, $T_{e,\parallel}$. Accordingly, $\gamma$ becomes about 1.75 by multiplying 2 by the coefficient of the relation $\partial(e\phi)/\partial s = 1.17 \partial T_{e,\parallel}/\partial s$ obtained by the experimental values. Considering that the estimated $\gamma$ exceeds 5/3 in the relation between the electron density and $e\phi$, it was concluded that the instability may play a crucial role in the plasma dynamics of the magnetic nozzle, increasing the effective collision frequency and cross-field transport. In the study, detailed experimental and theoretical support for instability and cross-field transport in the MN were not provided. However, in recent years, relevant research has been conducted on a topic related to (Hepner and Jorns., 2021) or independent of thermodynamics (Singh *et al*., 2013; Hepner *et al*., 2029), and it was revealed that instabilities can increase turbulent collision frequency of electrons. The enhanced cross-field transport and modification of the EVDF accompanied by the instability is believed to affect the electron cooling physics in the divergent magnetic field. Therefore, we expect that an in-depth discussion can reinforce the contents of the scattered data in the nozzle throat and the far-field region.

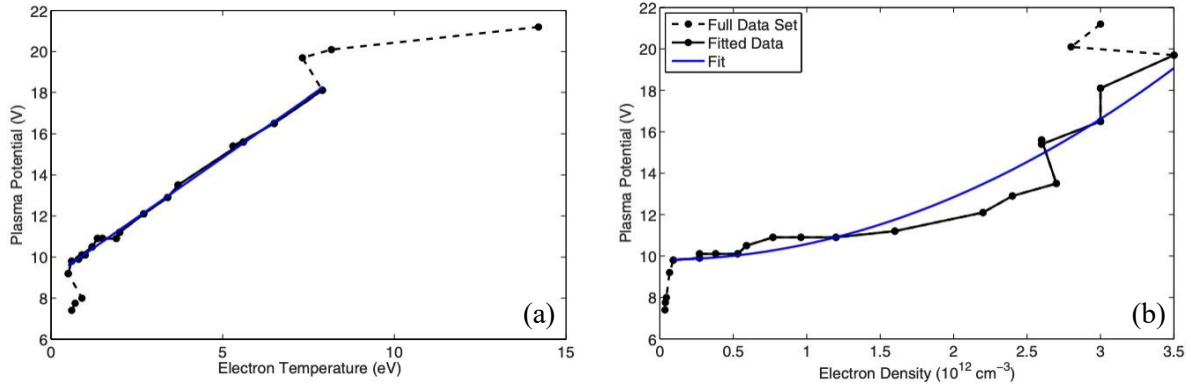

FIG. 2. Plasma potential $e\phi$ versus (a) parallel component of electron temperature $T_{e,\parallel}$ and (b) density $n_e$ in the magnetic nozzle. The blue line in (a) and (b) is $e\phi \propto T_{e,\parallel}$ and $e\phi \propto n_e^2$ fit, respectively. The dashed lines connecting scattered data points were not included in the fitted data set. Reproduced with permission from Ref. (Sheehan *et al.*, 2014). Copyright 2014 IOP Publishing.

The thermodynamic studies conducted in MNs and Hall thrusters have in common that they try to reveal "the relationship between the heat flux of electrons at the exit of the plasma source and the ion energy" through the combination of the polytropic equation and the momentum equation. Using a similar approach, Lafleur et al. (Lafleur *et al.*, 2015) suggested the relationship of maximum ion energy to electron temperature at thruster exit plane and the polytropic index. The experiment observed changes in electron temperature at the nozzle throat and ion energy in the expansion region when the mass flow rate and magnetic field strength were changed, and then the polytropic index is estimated.

The experiments were conducted for three electromagnet currents to determine the relationship between magnetic field, ion acceleration, and electron temperature: case A, No magnetic field; case B, moderate magnetic field (with the ECR condition located at the thruster back wall); case C, high magnetic field (with the ECR condition located in the center of the thruster). In the null field condition, surface wave absorption is the dominant heating mechanism in the discharge.

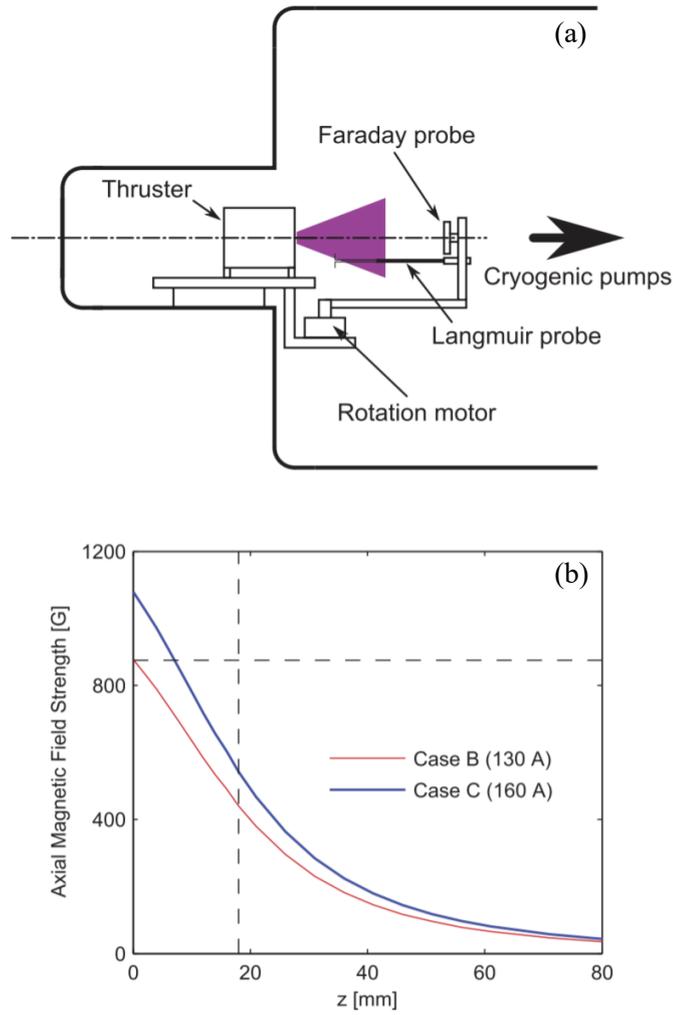

FIG. 3. (a) Schematic of electron cyclotron resonance (ECR) plasma thruster and diagnostics apparatus inside the space simulation chamber, and (b) axial profile of the magnetic field strength of the axial component for two cases B and C. The horizontal dashed line denotes the magnetic field strength 0f 875 G at which the ECR is expected to occur. Reproduced with permission from Ref. (Lafleur *et al.*, 2015). Copyright 2015 IOP Publishing.

    As the magnetic field was strengthened, the electron temperature increased while the ion energy did not show distinctive changes [Fig. 4], indicating that the magnetic field does not result in additional ion acceleration in the downstream region. The cooling rate in the electron temperature was larger than that of the ion energy with increasing the magnetic field strength, inferring the proportional relationship between polytropic index and magnetic field strength (see Fig. 5 and equation in the caption). Accordingly, this result provides a perspective that the high cooling rate of electron temperature far downstream is not directly related to ion acceleration.

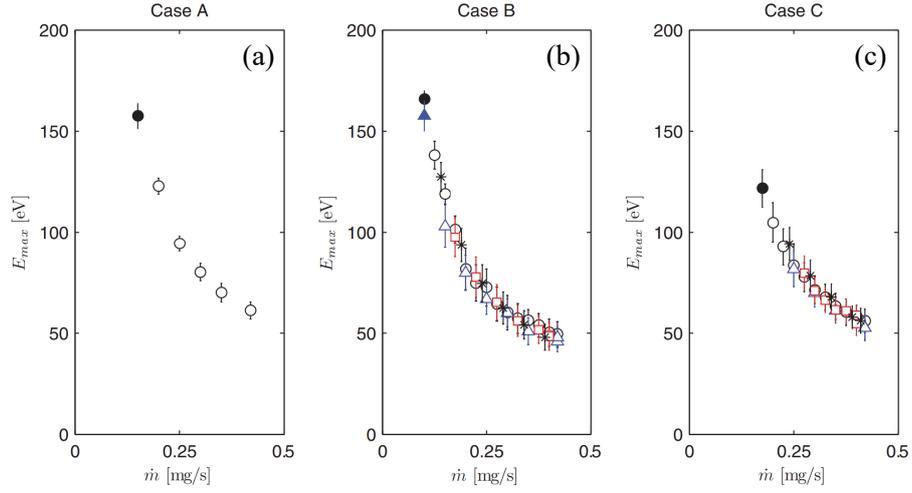

FIG. 4. Measured maximum ion energy, $E_{max}$, as a function of mass flow rate, $\dot{m}$, for (a) case A, (b) case B, and (c) case C. The different symbols and colors of data show multiple sets of experiments. Reproduced with permission from Ref. (Lafleur *et al.*, 2015). Copyright 2015 IOP Publishing.

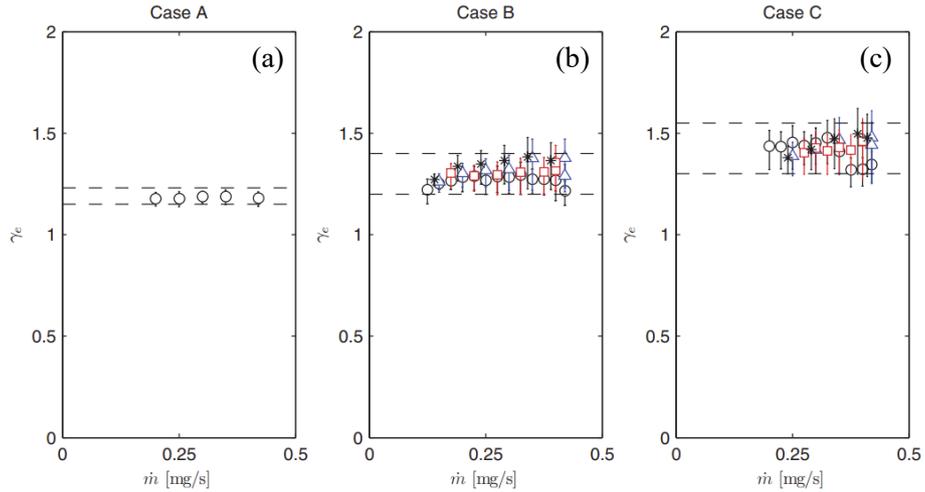

FIG. 5. Calculated polytropic index of electrons, $\gamma_e$, as a function of mass flow rate, $\dot{m}$, for (a) case A, (b) case B, and (c) case C. The polytropic index is estimated by using the ratio of the maximum ion energy, $E_{max}$, to the upstream electron temperature, $T_{e0}$, $E_{max}/T_{e0}$, as follows; $E_{max}/T_{e0} = 0.5 + \gamma_e/(\gamma_e - 1)$. The different symbol and color of data show multiple sets of experiments. The horizontal dashed lines mark the lower and upper limits of the experimental values. Reproduced with permission from Ref. (Lafleur *et al.*, 2015). Copyright 2015 IOP Publishing.

As those authors stated, there is room for improvement in the following matters related to measurement in the identification of the electron thermodynamic state. First, as revealed in recent studies, the polytropic index is a value that varies along a divergent magnetic field line (Correyero *et al.*, 2019; Kim *et al.*, 2018). From this point of view, the absence of measured data at the nozzle throat where the "highest rate of electron cooling" is predicted (positions inside more than 8 cm) suggests the possibility that the (global) measured polytropic index is underestimated. The change in the strength of the magnetic field, which is an experimental manipulation variable, may cause a change in the position of the maximum ion energy; however, in this study, the position of the Faraday probe that serves as retarding field energy analyzer is fixed at 27 cm. In other words, it can be predicted that the magnetic expansion is not yet completed. The anisotropy of the electron temperature is expected to be strong due to the inherent electron heating mechanism of the ECR source, and it should not be overlooked in the future study. Nevertheless, this study is meaningful in that plasma variables are measured and analyzed from a kinetic perspective.

Experiments conducted in the helicon plasma source (Chi-Kung reactor) reported a different thermodynamic state of electrons from previous studies (Zhang *et al*., 2016a). Their logic was based on non-local electron kinetics in the nearly collisionless regime, which focused on the spatial change of the electron energy probability function (EEPF or *eepf*) following the generalized Boltzmann relation.

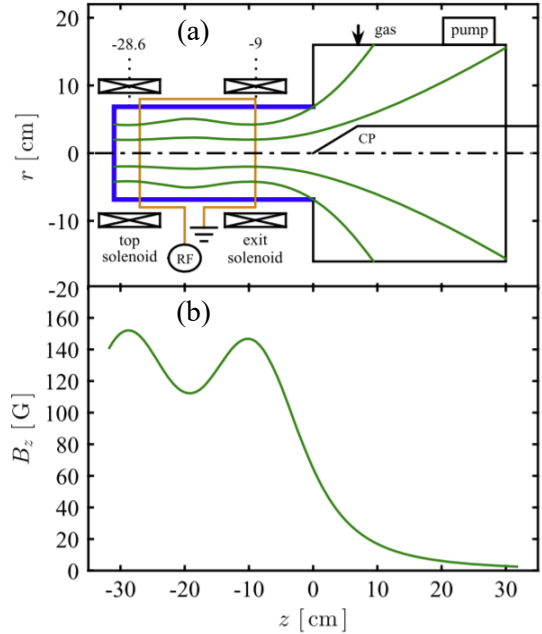

FIG. 6. (a) Schematic of Chi-Kung, the helicon plasma reactor, showing the major components, diagnostics probes, and magnetic field lines. (b) Magnetic flux, $B_z$, on the central axis. Reproduced with permission from Ref. (Zhang *et al*., 2016a). Copyright 2016 APS Publishing.

It was argued that the thermodynamics of electrons in a divergent magnetic field is governed by the non-local EEPF in which the total energy is conserved, and therefore the shape of EEPFs is identical along the axial direction except for the cutting of the low-energy electrons. In the study, EEPFs have a convex structure (Druyvesteyn-like distribution), and the calculated effective electron temperature (averaged electron energy) decreases along the axial direction. Accordingly, the electron system does not show dramatic cooling and has a polytropic index of 1.17 closer to the isothermal value.

Then, the electrons transfer their enthalpy into electric potential energy during the magnetic expansion, verifying an adiabatic process without thermal conduction into the system.

This logic indicates that the polytropic index is dependent on the shape of the non-Maxwellian EEPFs under the condition that the non-local kinetics is dominant (Boswell *et al.*, 2015). For instance, when the non-local electron kinetics dominates the magnetic nozzle system (total electron energy is conserved) and the shape of EEPFs is concave (bi-Maxwellian-like distribution) with the existence of high-energy groups, the decrease in the electric potential in the axial direction acts as a barrier to the low-energy groups. Thus, the low-energy electrons that cannot overcome the plasma potential decrease in the axial direction, and only electrons with high kinetic energy can reach the far-field. In this case, the effective electron temperature at the far-field region is higher than that of the nozzle throat under the identical electric potential structure, indicating that the polytropic index can be less than unity as analyzed by Zhang *et al* (Zhang *et al*., 2016b). After all, they concluded that there is a fundamental difference in interpreting the thermodynamics of particles in non-local and local thermodynamics equilibrium, and the polytropic index closer to unity is not a result of the heat conduction along the divergent magnetic field, but rather the result of the non-local property of electrons along the divergent magnetic field.

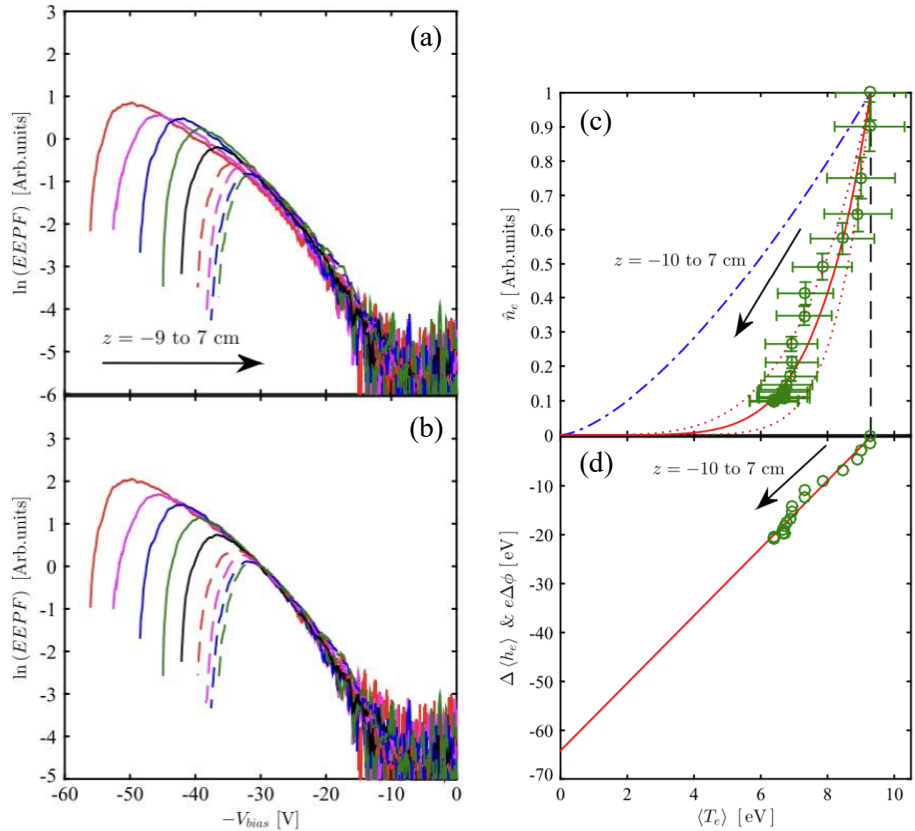

FIG. 7. (a) The logarithm of electron energy probability functions (EEPFs) as a function additive inverse of bias voltage on the single Langmuir probe, $-V_{bias}$ at each 2 cm from the axial location $z = -9$ to 7 cm. (b) EEFPs normalized at $-V_{bias} = -30$ V. The solid curves and dash-dotted curves represent the measured EEPFs in the plasma source ($z < 0$ cm) and diffusion chamber ($z > 0$ cm), respectively. (c) Correlation data between normalized electron density, $\hat{n}_e$, and effective electron temperature, $\langle T_e \rangle$. The polytropic curve with an index of 1.17 is plotted as a red solid curve. The upper and lower limit curves around the polytropic curve given as two dotted lines are obtained by fitting the experimental parameters. The dash-dotted line (red) and dashed line (black) represent the processes with a polytropic index of 5/3 and unity, respectively. (d) Relative electron enthalpy, $\Delta \langle h_e \rangle$ (solid line) and relative plasm potential, $\Delta \phi$ (open circles) as a function of $\langle T_e \rangle$. The electrons transfer their effective enthalpy into potential energy during the plasma expansion. See Ref. (Zhang *et al.*, 2016a) for a detailed explanation. Reproduced with permission from Ref. (Zhang *et al.*, 2016a). Copyright 2016 APS Publishing.

We carefully highlight the design factors of the MN (e.g., antenna, and magnetic field structure) in the reason for the difference in thermodynamic analysis of each group. Unlike the MN set-up of other research groups, the convergent magnetic field line is not clearly observed in the MN with RF source (Takahashi *et al.*, 2017a). Eventually, the localized wave heating of electrons near the antenna and its transport along the magnetic field line keeps the electron temperature at the center of the source radius lower than the peripheral radius edge region. As a result, the electron temperature at the radial center at the nozzle throat has a low electron temperature compared to other streamlines of the magnetic field at the same axial location [Fig. 8(a)]. Indeed, the electron temperature of the radial center at the nozzle throat is already closer to that of the outer streamline in the middle of the diffusion chamber, which departs from the plasma source. This phenomenon is also observed in the plasma parameters measured in the ICP nozzle with a double-turn antenna with a large-volume expansion chamber (argon 0.8 mTorr) [Fig. 8(b–d)].

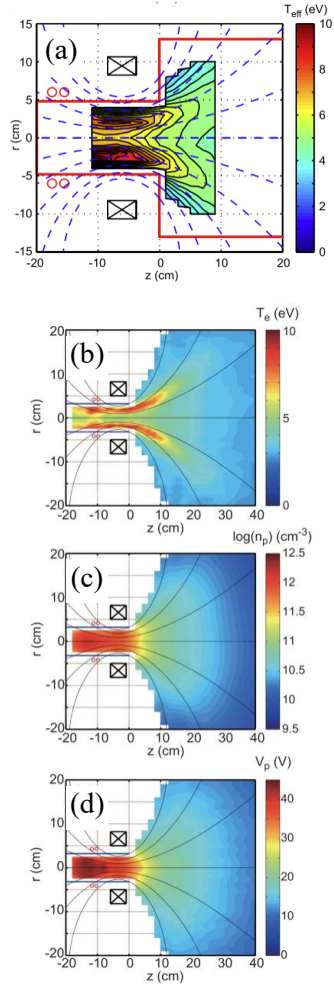

FIG. 8. 2D profiles of the electron temperature, $T_e$, measured in (a) small (26 cm diameter and 30 cm long) and (b) large (60 cm long and 140 cm long) diffusion chamber, respectively. A Pyrex source tube of 6.4 cm inner diameter and 20 cm long and 9.4 cm inner diameter and 20 cm long is immersed in each small and large diffusion chamber, respectively. (c) the logarithm of the plasma density, $log\ n_p$, and (d) the plasma potential, $V_p$, for the set-up of large diffusion chamber are depicted. Compared to the large radial gradient of electron temperature, the relatively uniform density and plasma potential in the radial direction of the source region are impressive. Reproduced with permission from Ref. (Takahashi *et al.*, 2017a). Copyright 2017 AIP Publishing.

A polytropic index close to 1 regardless of the magnetic field strength and structure was observed in an experiment where the driving pressure was about 10 times lower [Fig. 9] (Little and Choueiri, 2016). Interestingly, it is noticeable that the spatial change of ion energy distribution function (IEDF) is dependent on the magnetic field strength and structure [Fig. 9]. The measured IEDF shows that the ion acceleration slows down in the far-field region as the magnet current increases (the ion energy has a maximum of about 40 cm at a high magnetic current).

Based on the results of other research groups that the polytropic index can be a function of space, the following analysis is possible. In the linear regression of electron temperature vs electron density, the re-calculated polytropic index of the nozzle throat (8 data in the upper right from nozzle throat to 18 cm) is 1.23, and from thereafter to 30 cm, it has 1.01. When the fitted data set is further reduced (nozzle throat to 7.5 cm), the calculated polytropic index is approximately 1.43, approaching the adiabatic value of 5/3. Although this simple approach has the limitation of providing only phenomenological analysis, the change in the spatial electron cooling rate is a factor to be understood in improving the fundamental understanding of electron thermodynamics and improving the efficiency of MN devices.

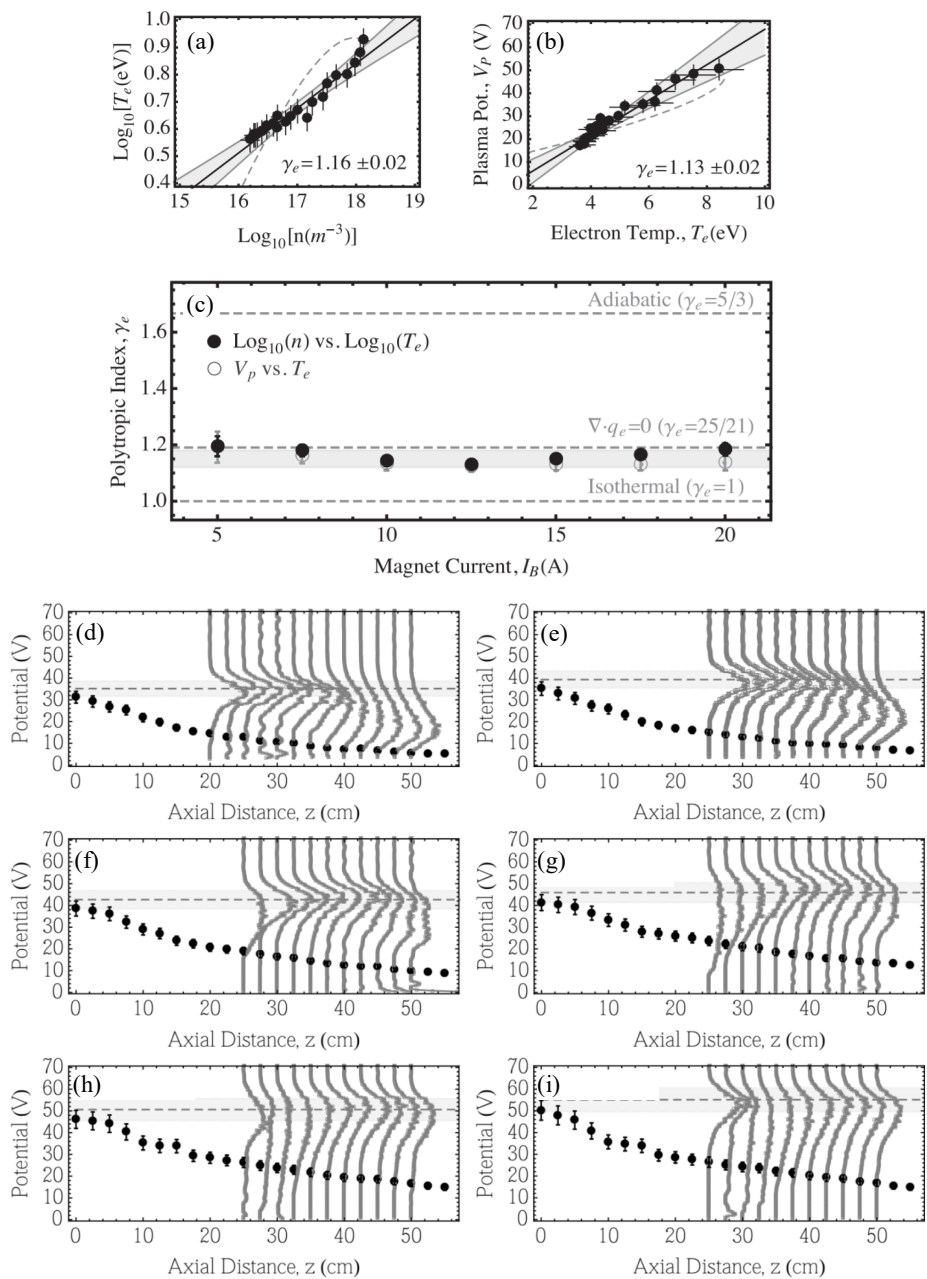

FIG. 9. Estimation of the polytropic index, $\gamma_e$, with experimental measurements of electron temperature, $T_e$, electron density, $n$, and plasma potential, $V_p$, with the relation of (a) $\log T_e$ versus $\log n$, and (b) $V_p$ versus $T_e$. The solution to the quasi-1D model is also shown (dashed line). The polytropic index is determined using the method of least squares (line). (c) the dependency of $\gamma_e$ on the magnet current, $I_B$, is shown (c). The axial evolution of the ion energy distribution function and $V_p$ with varying $I_B$ of (d) 5.0 A, (e) 7.5 A, (f) 10.0 A, (g) 12.5 A, (h) 15.0 A, and (i) 17.5 A. Reproduced with permission from Ref. (Little and Choueiri, 2016). Copyright 2016 APS Publishing.

## B. Effect of trapped electrons

Previous studies have defined the thermodynamic state of electrons by considering all electrons as a single system. As will be explained in Section IV, kinetic models of the plasma expansion (Martinez *et al*., 2015; Sanchez-Arriaga *et al*., 2015; Merino *et al*., 2021) in the MN suggest the existence of three electron subpopulations, occupying different regions of phase space: (1) free electrons coming from the source, and with enough energy to escape to infinity (these electrons are in charge of neutralizing the ion current emanating from the device), (2) reflected electrons coming from the source, and with insufficient energy to escape downstream, and (3) trapped electrons, whose magnetic moment and energy allow them to exist in an intermediate part of the MN, but whose orbits do not connect to the plasma source not to infinity. The studies dealt with in this section subdivide electrons into groups with different thermodynamic states. Such an attempt provides an essential answer to the question of what the thermodynamic state of electrons is in a magnetically expanding plasma.

Takahashi et al. investigate the thermodynamic state of electrons through a completely different experimental device from previous studies (Takahashi *et al*., 2018). The newly designed device succeeded in realizing an    electric-field-free system by excluding the electric field in the axial direction [Fig. 10]. While this differs substantially from the conditions in a MN, the device can control the plasma potential value and gradient in the axial direction, and consequently, the interaction between magnetic field and electrons can be explored under experimental conditions in which the effect of an axial electric field is completely excluded [Fig. 11].

When the potential difference is close to zero and the change in the axial direction is negligible, the electron temperature is rapidly cooled along the magnetic field, and the measured polytropic index is greater than 1.4, approaching the adiabatic value of 5/3 [Figs. 11 and 12]. On the other hand, when the potential difference is large and an electric field in the axial direction is formed, like a general MN, the thermodynamic state of electrons is close to isothermal. The study suggested that when the generation of trapped electrons by the electric field is suppressed, the electron system can work on the magnetic field alone, and the Lorentz force generated by the non-uniformity of the radial plasma density acts on the expanding magnetic field to form an ideal gas that expands adiabatically [Fig. 11]. It causes a decrease in the internal energy of the working electron. This means that the classical laws of thermodynamics can be extended to the expansion of a collision-free electron gas in a MN.

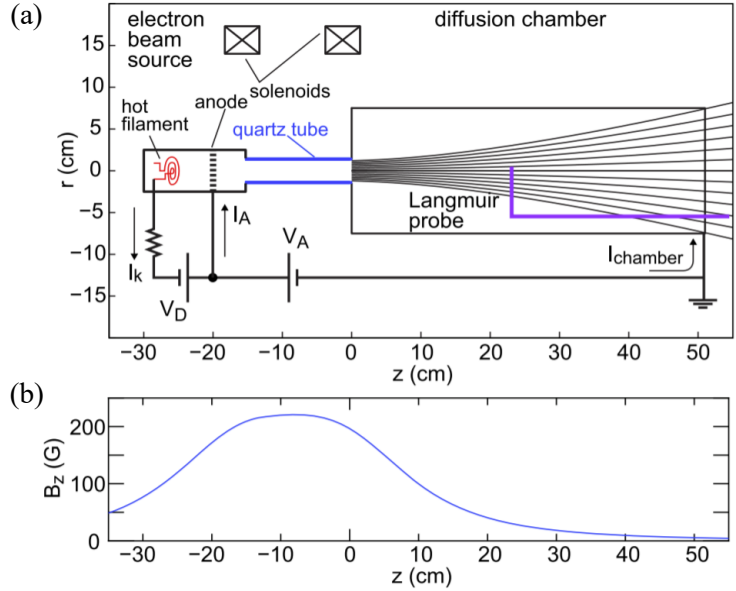

FIG. 10. (a) Schematic of the experimental setup. (b)Axial profile of the magnetic field on axis. In the specially designed setup, the plasma potential is mainly governed by the anode potential, which is an intrinsic characteristic of the DC plasma sources. Reproduced with permission from Ref. (Takahashi *et al*., 2018). Copyright 2018 APS Publishing.

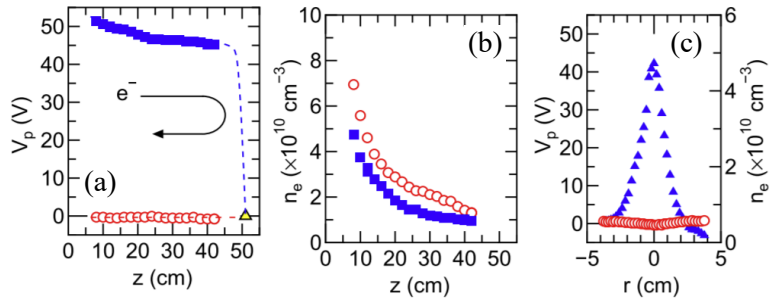

FIG. 11. (a) Axial profile of plasma potential, $V_p$, for anode potential, $V_A$, of 60 V (filled squares) and 0 V (open circles). A triangle shows the grounded wall potential at the axial location, $z$, of 51 cm. (b) Axial profile of electron density, $n_e$, for the same values of $V_A$. (c) Radial profile of $V_p$ (open circles) and, $n_e$ (filled triangles) measured at $z = 10$ cm for $V_A = 0$ V. Reproduced with permission from Ref. (Takahashi *et al*., 2018). Copyright 2018 APS Publishing.

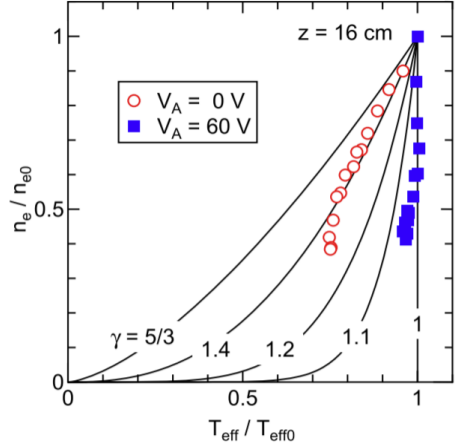

FIG. 12. Polytropic relation obtained from the measured electron energy probability functions, together with the theoretically calculated curves with spatially varying electron density, $n_e$, and effective electron temperature, $T_{eff}$, normalized by center value. Reproduced with permission from Ref. (Takahashi *et al.*, 2018). Copyright 2018 APS Publishing.

While Takahashi et al. study emphasized that only electron groups that undergo adiabatic expansion can be observed when the electric field is artificially removed, Kim et al. independently carried out experiments that observe the thermodynamic state of each electron group in a MN in the presence of an axial electric field (Kim *et al.*, 2018). They analyzed the electron thermodynamics under the perspective that a magnetic mirror formed by the combination of magnetic field and self-generated electric field can create trapped electron motion. A double-sided planar Langmuir probe is used to selectively collect electron groups in the expansion region of the MN device.

The presence of isothermally behaving electrons separates the MN system into two regions with different thermodynamic properties [Fig 13]. One is an adiabatic region located near the nozzle throat and the other is an isothermal region located downstream. This region separation effect is maximized when the strength of the magnetic field is increased. At high magnetic field strength, an abrupt change in effective electron temperature is observed at the nozzle throat by the front side of the probe (downward probe), and the calculated polytropic index is closer to 5/3. On the other hand, when the strength of the magnetic field is weak, the decrease rate of electron temperature becomes lower, and the polytropic index calculated by measured EEPFs by the downward probe has a value closer to unity in the entire area of the nozzle. Interestingly, the upward probe (back probe) only collects non-locally behaving low-energy electrons showing the isothermal polytropic index.

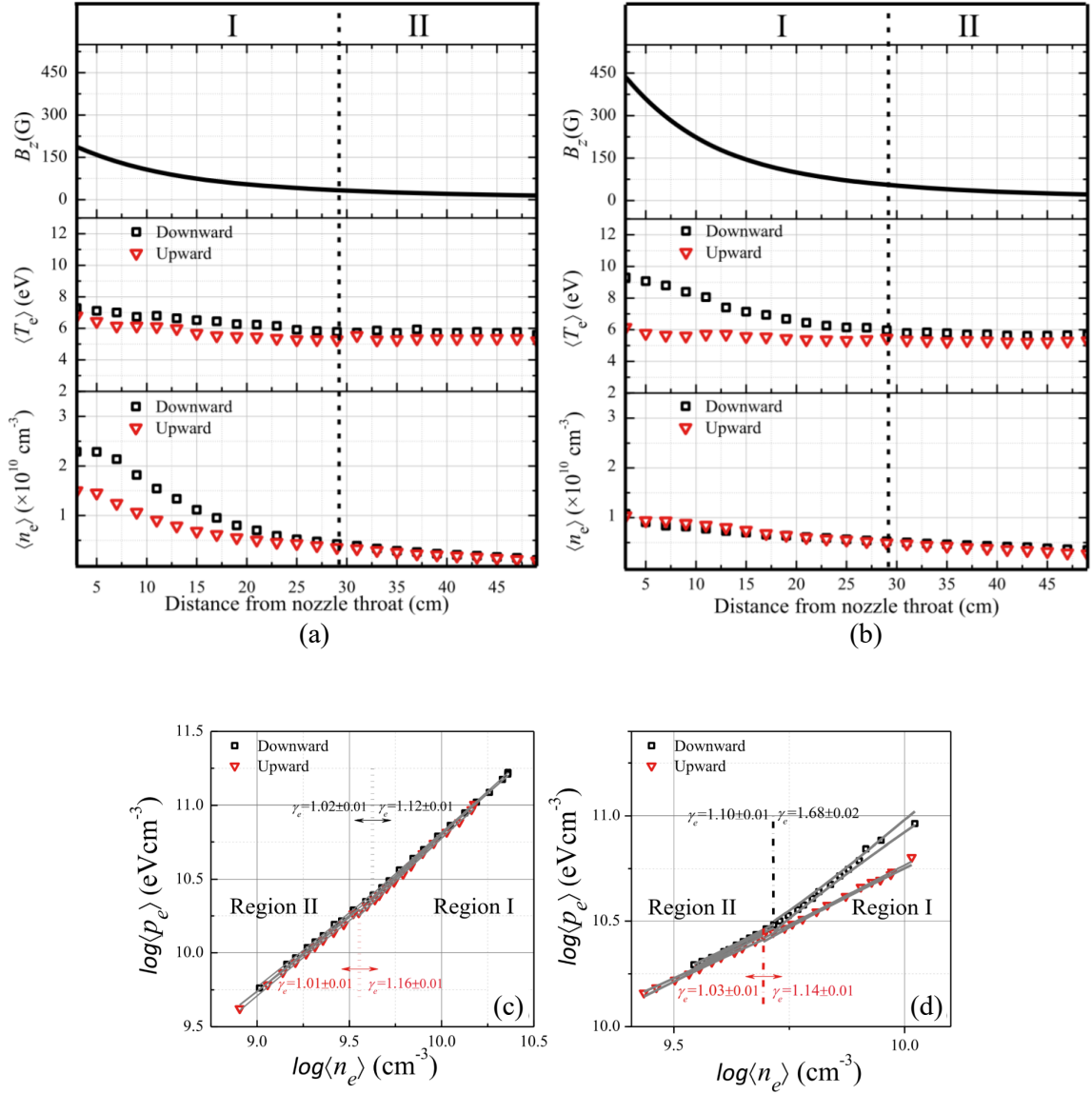

FIG. 13. Axial profile of magnetic flux, $B_z$, and electron parameters measured by the front probe (open squares) and back probe (open triangles) from 3 to 49 cm from the nozzle throat at (a) 50 A, and (b) 200 A of electromagnet current: effective electron temperature, $\langle T_e \rangle$, electron density $\langle n_e \rangle$. Log-log relationship between the effective electron pressure $\langle p_e \rangle$ and $\langle n_e \rangle$ averaged over 1D electron energy probability functions obtained at 2 cm intervals from 3 to 49 cm from the nozzle throat. The polytropic index of the MN system is determined by a combination of thermodynamic properties of isothermal and adiabatic electron groups, showing (c) the isothermal behavior at 50 A, and (d) the coexistence of adiabatic and isothermal groups near the nozzle throat and its evolution into the isothermal at the far-field region at 200 A. Reproduced with permission from Ref. (Kim et al., 2018). Copyright 2018 IOP Publishing.

The change in the thermodynamic properties of electrons varying with the strength of the magnetic field can be explained by the spatial formation of the maximum magnetic moment well. As the magnetic field strength increases, the bounce region of electrons (maximum magnetic moment well) moves to the far region of the MN. In other words, a polytropic index close to 5/3 is regarded as the result of a shift in the bounce region where the cooled electrons stagnate [Fig. 14].

Importantly, the studies of Kim and Takahashi provided experimental consensus on the different polytropic indexes found in each study group. The spatial distribution of two-electron groups with different thermodynamic states determines the polytropic index, and the properties of electric and magnetic fields possessed by devices in each group have a polytropic index ranging from 1 to 5/3. Ultimately, this study demonstrates that the thermodynamic properties of a magnetically expanding plasma should not be directly related to the MN efficiency.

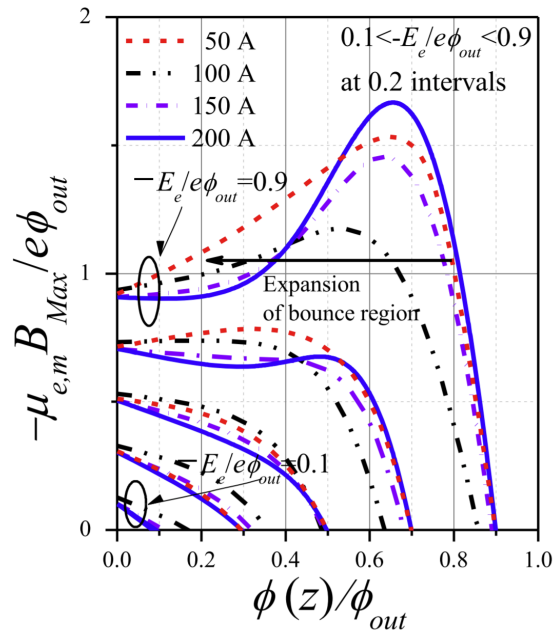

FIG. 14. For electrons, the local maximum magnetic moment $\mu_{e,m}(z, E_e)$ with total energy $E_e$ is expressed as follows: $\mu_{e,m}(z, E_e) = (E_e + e\phi(z))/B_z$. The local maximum magnetic moment have minimum and maximum values at points, which eventually clarify the bounce motion of electrons in a magnetic nozzle system. The confined electrons having energy below the total potential drop then bounce back (reflected and trapped electrons) and forth (trapped electrons) in the bounce region. The graph shows the normalized local maximum magnetic moment $-\mu_{e,m}B_{Max}/e\phi_{out}$ versus normalized plasma potential drop $\phi(z)/\phi_{out}$ for various normalized electron energies $E_e/e\phi_{out}$ at nozzle current increasing from 50 to 200 A. Reproduced with permission from Ref. (Kim *et al.*, 2018). Copyright 2018 IOP Publishing.

Previous thermodynamic studies have been limited to static observation of plasma that has reached steady-state, and time-dependent kinetic analysis successfully identified a series of electron cooling processes in a MN (Kim *et al.*, 2019). By controlling the diffusion of the source plasma into the expansion region using a mesh grid installed at the boundary of the source and expansion region [Fig.

15], a series of electron cooling, generation of the ambipolar electric field, and production of trapped electrons could be observed. The gradually accumulated electrons change the low energy of the EEPFs [Fig. 16]. This accumulation of trapped electrons reduces the degree of cooling of the system, and the electric field initially generated by the adiabatic expansion in the downstream region disappears due to disconnection from the source.

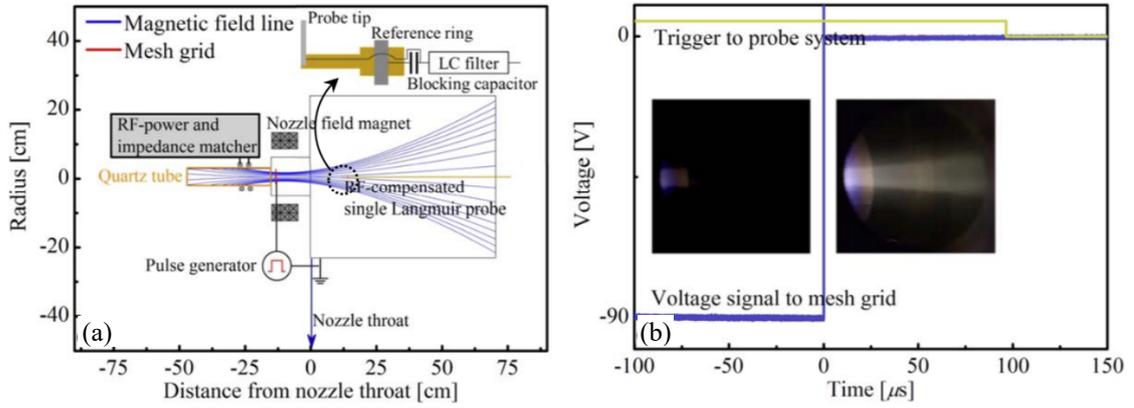

FIG. 15. Schematic diagram of (a) the magnetic nozzle device driven by the inductively coupled plasma source and the divergent magnetic field configuration. An axially movable RF-compensated single Langmuir probe is located at the expansion region. To observe a series of electron expansion, a mesh grid is installed at −13 cm from the expansion chamber throat, and (b) the voltage signal to the mesh grid and trigger to probe system (the internal images, which were taken under the steady-state condition at each voltage, are inserted to aid the understanding of the experiment). Reproduced with permission from Ref. [25]. Copyright 2019 IOP Publishing.

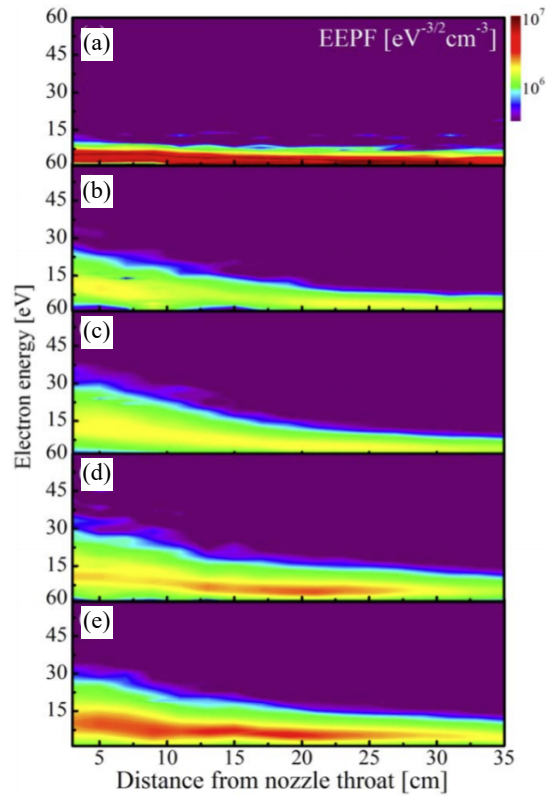

FIG. 16. Time evolution of the electron energy probability functions (EEPFs) at (a) −0.5 μs, (b) 3.0 μs, (c) 7.5 μs, (d) 25 μs, and (e) 95 μs relative to the beginning of the pulse rise time (0 μs). Reproduced with permission from Ref. (Kim *et al.*, 2019). Copyright 2019 IOP Publishing.

The log-log relationship of data shows that the adiabatic process dominates the electron thermodynamics near the nozzle throat at all moments [Fig. 17]. That is, the thermodynamic states of the electrons near the nozzle throat are maintained over time. Up to 3.0 ms, a slope of the log-log plot (i.e., the polytropic index) does not change during the entire expansion region. In contrast, a temporal variation of the slope is observed in the far-field region. The evaluated polytropic index is closer to unity as it approaches the downstream, indicating that the gradually accumulated trapped electrons in the downstream region behave to preserve the thermal energy with time. This study suggests the fundamental cause of the spatially varying polytropic index, emphasizing that the consideration of the trapped electrons is an essential factor for understanding the characteristics of a magnetically expanding plasma.

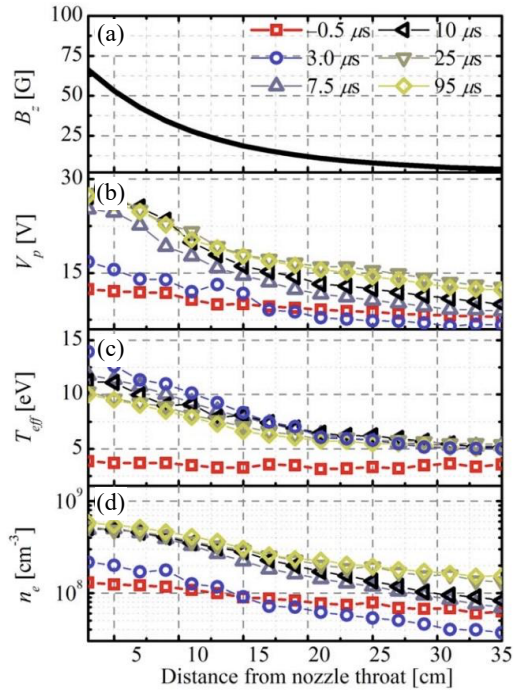

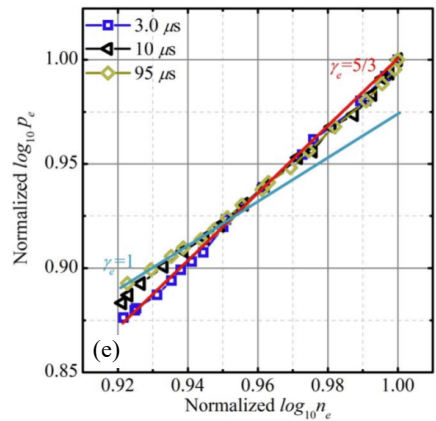

FIG. 17. Axial profile of (a) magnetic flux, $B_z$, and that of electron parameters from 3 to 35 cm from nozzle throat over time: (b) the plasma potential, $V_p$, (c) the effective electron temperature, $T_{eff}$, and (d) the electron density, $n_e$. (e) Log–log relationship between the effective electron pressure $p_e$ and $n_e$. Polytropic curves with an exponent of 5/3 (solid red) and unity (solid blue-green) represent the adiabatic and isothermal process, respectively. Reproduced with permission from Ref. (Kim *et al*., 2019). Copyright 2019 IOP Publishing.

## C. Changes in the degree of freedom

Generally, in the experiments and modeling performed on the MN device, the electron degrees of freedom were set to 1 and 2 in the parallel (axial motion) and perpendicular (radial and azimuthal motion) to magnetic field line in cylindrical coordinate, respectively. Accordingly, the adiabatic limit of the polytropic index would correspond to 5/3 in the MN.

Interestingly, it was found that the control of the degree of freedom in the MN device was made possible through the reduction of cross-field transport via radial electric field (Kim *et al*., 2021a; Takahashi *et al*. 2018). The strengthening of the radial electric field was achieved through the increase of the magnetic field strength, and it was eventually proved that the reduced degree of freedom increases the electron cooling rate (polytropic index). Ultimately, the essence of the relationship between degree of freedom and electron thermodynamics can be understood by controlling the following variables: 1) strengthening the radial electric field to restrict the cross-field transport of plasma 2) eliminating the axial electric field to prevent the electrons from trapped motion along magnetic field line. To minimize the axial electric field and maximize the radial electric field, DC filament plasma source is installed in the source region where the plasma potential is determined by the anode potential; the grounded chamber wall was designated as anode; thus, the plasma potential is closer to zero (Kim *et al*., 2021a).

Since the cross-sectional area of the expansion of beam-plasma (which is dominated by the size of filament in the source region) is excessively smaller than the expansion region, the radial electric field was generated in the expansion region. The aforementioned electric field formation is an intrinsic property of DC or indirectly heated cathode discharges that generate beam-plasma (Kim *et al*., 2021c).

The strength of the magnetic field was changed by controlling the current of the nozzle field coil, and the gradient scale length of the magnetic field was changed through an additional guiding coil to change the structure of the magnetic field [Fig. 18].

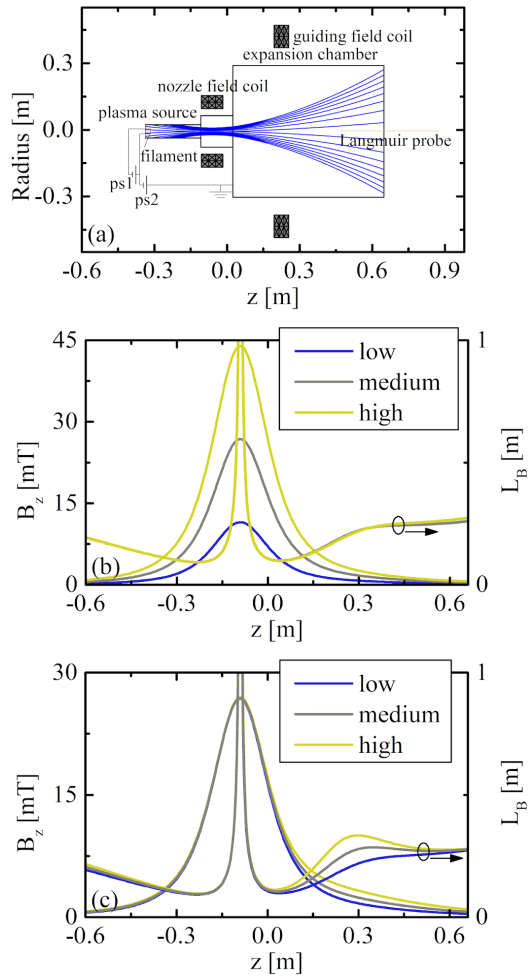

FIG. 18. Schematic diagram of the axial symmetric magnetic nozzle showing filament plasma source and the divergent magnetic field configuration with axially movable single Langmuir probe. (b) and (c) show magnetic field condition for strength, $B_z$, and structure variation $L_B = B_z/\nabla B_z$, respectively. Reproduced with permission from Ref. (Kim *et al*., 2021a). Copyright 2021 IOP Publishing.

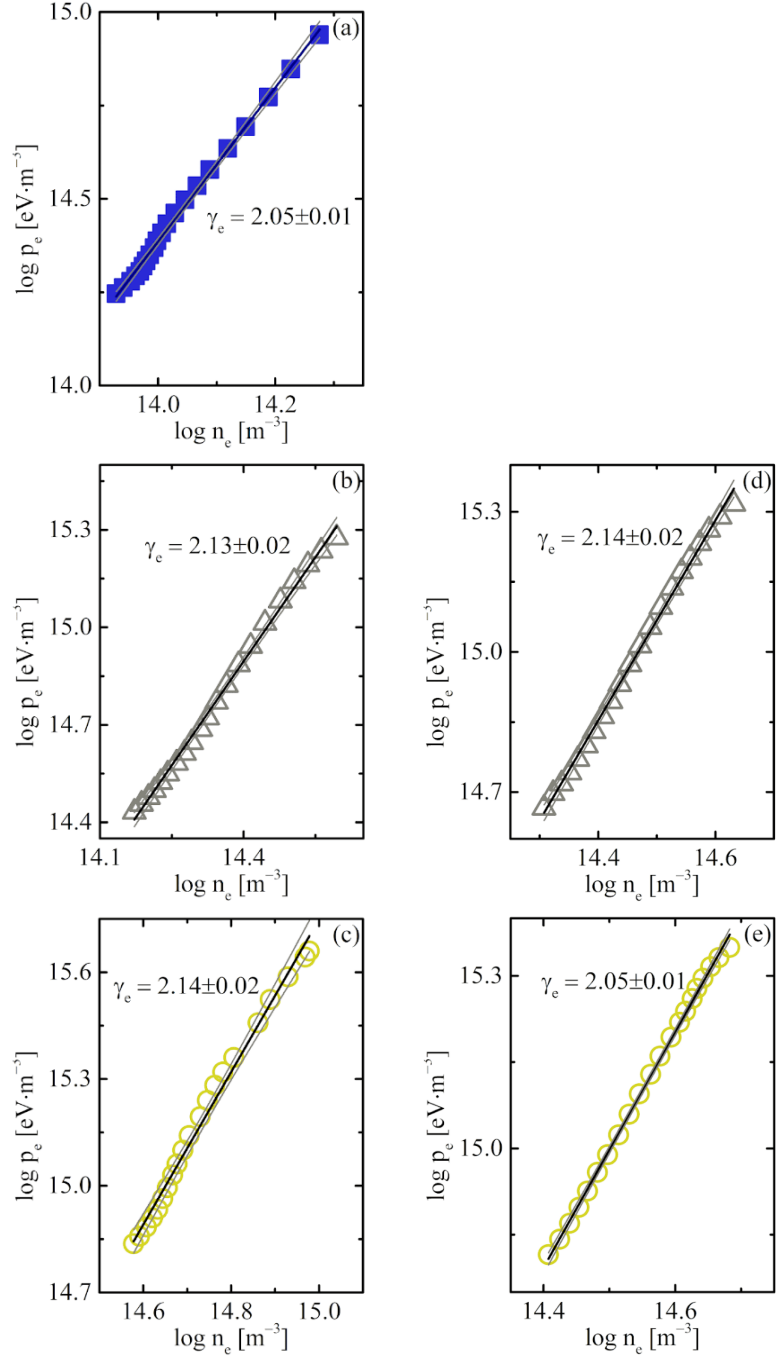

FIG. 19. Dependency of the polytropic index, $\gamma_e$, on magnetic field strength, $B_z$; (a) low, (b) medium, and (c) high $B_z$. The medium $B_z$ condition is assigned as low gradient scale length, $L_B$. (d) and (e) are results of medium and high $L_B$ structure, respectively. The polytropic index determined by log–log relationship between the electron pressure, $p_e$, and the electron density, $n_e$, averaged over the electron energy probability function. Reproduced with permission from Ref. (Kim *et al.*, 2021a). Copyright 2021 IOP Publishing.

Although the strength and structure of the magnetic field were changed, the plasma potential structure in the axial and radial direction was kept constant while the electron temperature and density gradient were different in each experimental condition (Kim *et al*., 2021a). Nevertheless, the fixed electric field strength regardless of the magnetic field properties ensures the invariance of the polytropic index closer to 2, indicating the reduced degree of freedom to 2 [Fig. 19].

The change of the polytropic index by the cross-field diffusion was proposed by earlier study of Takahashi *et al*. (Takahashi *et al*., 2020), but they did not introduce the concept of degree of freedom. Takahashi et al. pointed out the limitation in that studies on the investigation of electron thermodynamics in magnetic nozzle devices focused only on axial plasma variables and provided a new perspective based on the radial variation of plasma parameters. When the magnetic field strength is increased, both the magnetization of the electrons and the electric field confining the ions radially are enhanced; the polytropic index approaches the adiabatic value of 5/3 [Fig. 20]. The direction of the electric field was towards the radial center and the strength was sufficient to confine the ions, and this effect limited the cross-field transport of electrons. Based on these experimental results, they identified the dependence on the magnetic field strength and the thermodynamic state of electrons; the polytropic index becomes dependent on (and proportional to) the magnetic field strength.

After all, if the magnetic field strength becomes stronger in their experiments to completely limit the cross-field transport of electrons and ions in the radial direction, the polytropic index will be close to 2. That is, as in Kim's study, when a sufficient radial electric field is ensured, it is expected that the polytropic index of an adiabatic value is 2 due to the reduced degrees of freedom of electrons regardless of the strength and structure of the magnetic field.

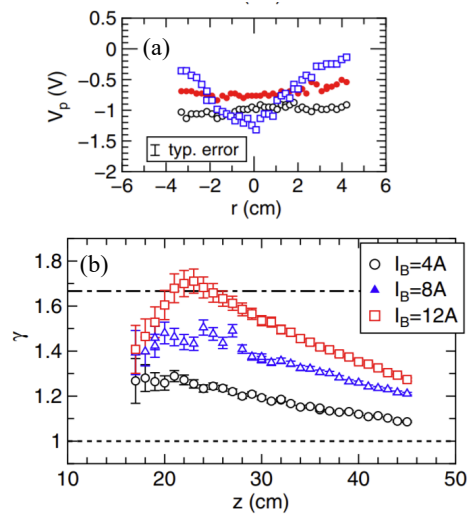

FIG. 20. (a) The radial profile of plasma potential, $V_p$, at the axial location, $z$, of 20 cm: 4 A (open circle), 7 A (filled circle), and 13 A (open square). (b) Typical axial profiles of the polytropic index, $\gamma$, calculated from the measured electron density and temperature. Reproduced with permission from Ref. (Takahashi *et al*., 2020). Copyright 2020 APS Publishing.

Finally, we discuss the evolution of EEPFs during the adiabatic process in a divergent magnetic field. Interestingly, the measured evolutions of EEPFs are close to the Maxwellian distribution function at the nozzle throat while the non-Maxwellian EEPFs were prominent in the far-field region of the MNs [Fig 21].

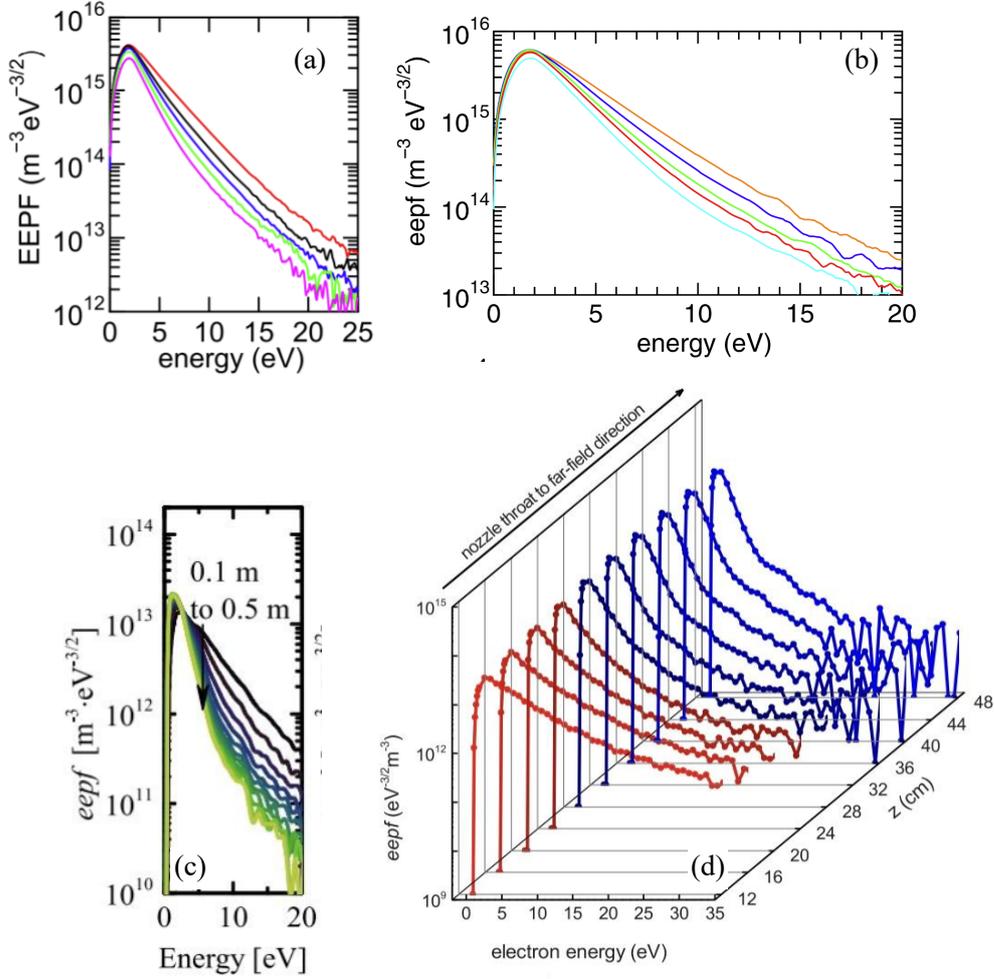

FIG. 21. Axial variation of electron energy probability function (EEPF or *eepf*) in electron kinetic energy scale. Axial location of (a) 15 (upper most curve) to 40 cm (downer most curve), (b) 16 (upper most curve) to 36 cm (downer most curve), (c) 10 (upper most curve) to 50 cm (downer most curve), and (d) 12 to 48 cm from the nozzle throat. Reproduced with permission from Ref. (Takahashi *et al*., 2018). Copyright 2018 APS Publishing. Reproduced with permission from Ref. (Takahashi *et al*., 2020). Copyright 2020 APS Publishing. Reproduced with permission from Ref. (Kim *et al*., 2021a). Copyright 2021 IOP Publishing. Reproduced with permission from Ref. (Kim *et al*., 2021b). Copyright 2021 APS Publishing.

The explanation for this phenomenon was clarified by the adoption of non-extensive thermodynamics (Kim *et al*., 2021b). In the MN device, the EEPFs can be fitted through the kappa function [Fig. 22]. Currently, it has been revealed that the cooling of electron temperature and the maintained kappa values along the expanding magnetic field represent a *reversible* and *adiabatic* process, respectively [Fig. 23]. That is, it indicates that the changes into non-Maxwellian EEPFs along the divergent magnetic field are an inevitable result of the thermodynamic process. The interpretation of the study can be extended to whether electrons with a non-Maxwellian distribution satisfy the laws of thermodynamics. By introducing the non-extensive statistical mechanics, they found an answer to the fundamental question of whether collisionless, magnetically expanding, non-equilibrium electrons satisfy the laws of thermodynamics via non-extensive Tsallis entropy.

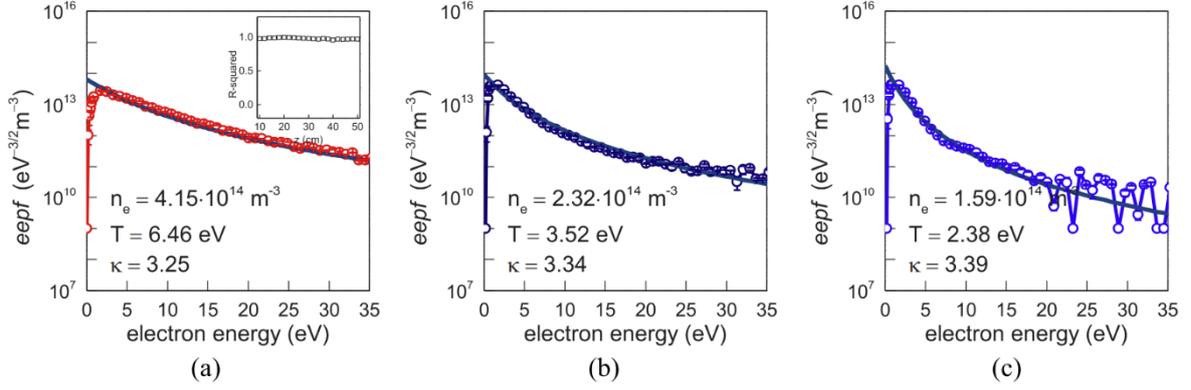

FIG. 22. Fitting of electron energy probability function (*eepf*) by the kappa distribution. The kappa distribution is a function of two independent parameters, temperature, $T$, and kappa, $\kappa$. (a) 12 cm, (b) 28 cm, and (c) 48 cm. The inset shows the R-squared values (the proportion of the variance of the fitted curve and the experimentally obtained *eepf*s in the range from 3 to 35 eV. Reproduced with permission from Ref. (Kim *et al*., 2021b). Copyright 2021 APS Publishing.

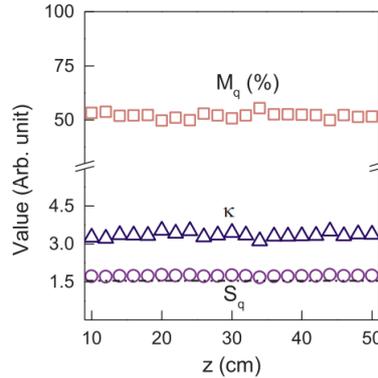

FIG. 23. Axial variation of $q$-metastability $M_q$, $\kappa$, and $S_q$. Quantifying the entropy enables discussion of the energy flow of the electrons in a magnetic nozzle. The non-extensive entropy $S_q$ in terms of $\kappa$ is given by $S_q(\kappa) = \kappa - \kappa^{\frac{1}{\frac{2}{3}\kappa+1}} \left[ \pi^{-\frac{3}{2}} \left( \kappa - \frac{3}{2} \right)^{\kappa - \frac{1}{2}} \frac{\Gamma(\kappa)}{\Gamma\left(\kappa - \frac{1}{2}\right)} \right]^{\frac{1}{\kappa + \frac{3}{2}}}$. The dashed line indicates the statistical

minimum of $\kappa$, 3/2. The kappa obtained along the axial direction is nearly constant at $3.35 \pm 0.05$. Thermodynamic distance of each stationary state from equilibrium through the $q$-metastability $M_q = 4[(q-1)/(q+1)]$, where the equilibrium is described by the classical equilibrium limit $M_q = 0$ for $q \to 1$ and the $q$-frozen state $M_q = 1$ for $q \to 5/3$, which is the state 100% away from equilibrium. The calculated $M_q$ (expressed as a percentage) for all axial positions is within $52 \pm 0.7\%$, implying invariance of the equilibrium state. Reproduced with permission from Ref. (Kim *et al.*, 2021b). Copyright 2021 APS Publishing.

In summary, various experimental studies to understand the cooling of electrons in magnetic nozzle devices have been summarized. The study of electron thermodynamics has been extended to considering the relationship between the trapping of electrons, the cross-field diffusion, and the degree of freedom of electrons and polytropic index. As the research that started with a general device to generate ion beams was subdivided into basic physical research using filament sources, it was possible to separate and observe electron groups with different thermodynamic properties, and finally suggest the following main points. The adiabatic expansion of electrons contributes to the formation of an electric field, which contributes to the creation of various groups of electrons, including trapped electrons. Therefore, in order to understand the physics of magnetic nozzle devices and to suggest engineering directions for performance improvement, it is essential to group electrons with distinct dynamic and thermodynamic characteristics in electric and magnetic fields.

## III. THEORETICAL APPROACH TO ELECTRON THERMODYNAMICS

This section presents existing models providing theoretical grounds to main physical phenomena observed in MN plasma expansions. In the first part (A), a two-fluid model is discussed (Ahedo and Merino, 2010; Merino and Ahedo, 2016), which has been successful in explaining the fundamentals of the operation of magnetic nozzles (ion acceleration, magnetic thrust generation and azimuthal electric currents (Ahedo and Merino, 2010), effect of collisions and electron inertia (Ahedo and Merino, 2012), formation of double layers in the presence of two-temperature electron distributions (Ahedo and Martinez-Sanchez, 2009; Ahedo 2011b; Merino and Ahedo, 2013), plasma detachment (Merino and Ahedo, 2014), effects of ion temperature (Merino and Ahedo, 2015), effect of the plasma-induced magnetic field (Merino and Ahedo, 2016), and contactless thrust vectoring (Merino and Ahedo, 2017)). The model uses a simple, empirical isothermal or polytropic closure relation for the electron species, and therefore it renounces to analyze the causes leading to electron cooling and temperature anisotropy which is left for the second part (B). That part covers recent developments on the kinetic modeling of electrons in the MN. First, a steady-state, collisionless model is discussed that makes manifest the existence of distinct electron subpopulations (free, reflected, and doubly-trapped electrons) (Martinez-Sanchez *et al.*, 2015; Ramos *et al.*, 2018; Ahedo *et al.*, 2020; Merino *et al.*, 2021). Then, a time-dependent model is reviewed that is able to recover the filling of the trapped electron phase space via the initial transient set up process of the plume (Sanchez-Arriaga *et al.*, 2018) and via collisions (Zhou *et al.*, 2021). Conclusions drawn from these models, as well as their limitations and pending work, are discussed too.

### A. Two-fluid framework of magnetized plasma expansions in space

In the following we restrict ourselves to an electron-driven, divergent, axisymmetric magnetic nozzle, in which electrons are warm and ions are cold, except where otherwise noted. Under the assumption the plasma in the MN is collisionless and quasineutral, and composed of fully-magnetized electrons and partially-magnetized, cold ions, the steady-state expansion is described by the following continuity and momentum equations:

$$\nabla \cdot (n\bm{u}_i) = 0, \tag{1}$$

$$m_i \bm{u}_i \cdot \nabla \bm{u}_i = -e\nabla\phi + e\bm{u}_i \times \bm{B}, \tag{2}$$

$$\nabla \cdot (n u_{\|e} \bm{1}_\|) = 0, \tag{3}$$

$$0 = -\nabla \cdot \bar{P}_e + en\nabla\phi - en u_{\theta e} B \bm{1}_\perp, \tag{4}$$

where the electron bulk velocity has been written as

$$\bm{u}_e = u_{\|e} \bm{1}_\| + u_{\theta e} \bm{1}_\theta, \qquad u_{\perp e} = 0, \tag{5}$$

and $\{\bm{1}_\|, \bm{1}_\perp, \bm{1}_\theta\}$ is a right-handed, magnetically-aligned vector basis, with $\bm{1}_\|$ and $\bm{1}_\perp$ in the meridian plane. All symbols above are conventional. Observe that equations Eqs. (4) and (5) retain zeroth-order

Larmor radius effects only, and in particular, Eq. (4) disregards electron inertia, which is negligible compared to ion inertia.

To close the fluid equation hierarchy at this level (i.e., without involving the energy equation), the pressure tensor $\bar{P}_e$ is assumed to be isotropic, so that $\nabla \cdot \bar{P}_e = \nabla p_e$, and moreover, a closure relation for the scalar pressure of the polytropic form

$$p_e \propto n^\gamma \qquad (6)$$

is imposed, with the polytropic coefficient $\gamma$ an empirical constant. Observe that $\gamma = 1$ yields the isothermal limit, while $\gamma = 5/3$ is the adiabatic value for electrons with three degrees of freedom.

This closure has the additional advantage that Eq. (4) can be integrated in the parallel direction, yielding:

$$H_e = \frac{\gamma}{\gamma-1} T_{e0} \left[ \left(\frac{n}{n_0}\right)^{\gamma-1} - 1 \right] - e\phi \qquad (7)$$

(for $\gamma \neq 1$), where $T_{e0}$ and $n_0$ are reference upstream values of the electron temperature and plasma density, respectively. The integration constant $H_e$ is magnetic line-dependent, and can vary across them. In fact, $H_e$ is fully determined from the conditions at the magnetic throat, which is the section of maximum magnetic field strength.

Then, taking the perpendicular projection of (Eq. (4)), we find

$$u_{\theta e} = -\frac{\mathbf{1}_\perp \cdot \nabla H_e}{eB}. \qquad (8)$$

which provides the azimuthal electron velocity along the MN given the field strength $B$ and the value of $H_e$ upstream.

These expressions can be used to eliminate $\phi$ in the ion equations (Eqs. (1) and (2)), which then become analogous to the Euler gasdynamics equations with the pressure provided by the electrons and extra source terms due to the magnetic force on the plasma,

$$\boldsymbol{F}_M = eB[(u_{\theta i} - u_{\theta e})\mathbf{1}_\perp - u_{\perp i}\mathbf{1}_\theta]. \qquad (9)$$

The hyperbolic differential equations for the supersonic ion flow can be solved for $\boldsymbol{u}_i$ and $n$ with common techniques (method of characteristics, finite volumes, discontinuous Galerkin, etc). Finally, the electron continuity equation in Eq. (3) can be used to compute $u_{\parallel e}$, as electron streamtubes coincide with magnetic streamtubes.

References (Ahedo and Merino, 2010; Merino and Ahedo, 2013) contain a detailed account of the dynamics of this system, including the ion acceleration and thrust generation mechanisms. The main driver of the expansion is the electron thermal energy. As in an unmagnetized plasma plume, the parallel thermal energy of electrons is converted to directed kinetic energy of ions thanks to the electrostatic potential $\phi$. This is referred to as ambipolar acceleration. The main advantage of the MN, however, is the following: the perpendicular electron thermal energy, which would be wasted in the absence of a guiding magnetic field, is converted to parallel electron thermal energy. This energy is then available for the continued acceleration of ions.

The force that converts the perpendicular to parallel energy is the magnetic force on the electron fluid,

$$F_{Me} = eBu_{\theta e}\mathbf{1}_\perp \equiv -(\mathbf{1}_\perp \cdot \nabla H_e)\mathbf{1}_\perp, \quad (10)$$

and is the largest term in Eq. (9). The reaction to this force is felt on the magnetic circuit that generates the MN; the parallel component of this reaction is termed (electron) *magnetic thrust*. Positive magnetic thrust results from electric currents in the plasma that induce a magnetic field which opposes the applied one (i.e., diamagnetic currents). In standard magnetic nozzles, the gradient $\nabla H_e$ at the throat decreases radially, and the electron azimuthal current is always and everywhere diamagnetic (i.e., thrust producing). This is so even if part of the azimuthal electron current downstream is due to the $\mathbf{E} \times \mathbf{B}$ drift, which in usual conditions is paramagnetic (i.e., drag producing).

The rest of the terms in Eq. (9), due to the ions, can be diamagnetic or paramagnetic. For initially non-rotating ions (i.e., zero ion swirl at the throat), the magnetic force on ions is zero initially, and rather small but paramagnetic downstream.

The magnetic force on electrons, $F_{Me}$, scales with $H_e$, which in turn scales with $T_{e0}$ and depends on the cooling rate $\gamma$. Consequently, so does the ion momentum gain and the magnetic thrust produced by the MN. Figure 24 shows the computed in-plane ion velocity in a MN with polytropic ($\gamma = 1.3$) and isothermal ($\gamma = 1$) electrons, where the differences are evident. It is therefore paramount to determine the thermodynamics of the electrons in the collisionless MN expansion, and in particular, the electron cooling, to evaluate the performance of the device, including the magnetic thrust.

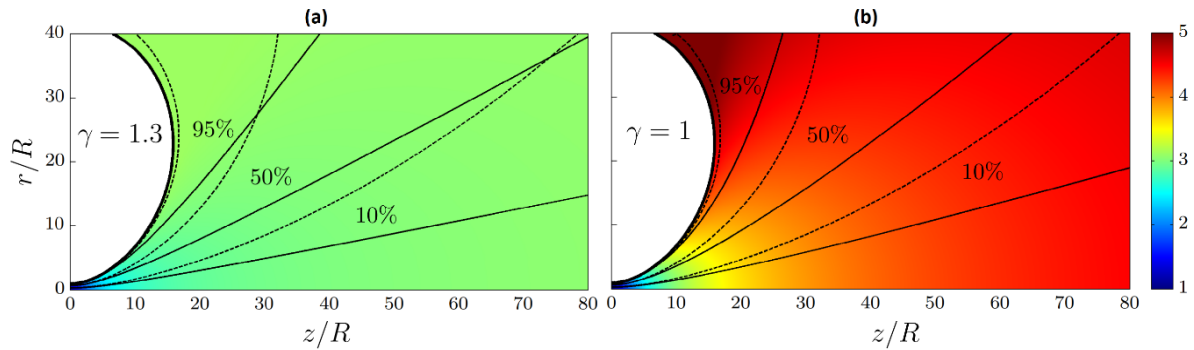

FIG. 24. Dimensionless in-plane ion velocity $u_i/\sqrt{m_i/T_{e0}}$ in a magnetic nozzle with polytropic ($\gamma = 1.3$, a) and isothermal (b) electrons. The ion velocity has been normalized using $T_{e0}$, the electron temperature at the origin. Dashed lines represent magnetic streamtubes; solid lines are ion streamtubes carrying a given percentage of the total ion flux as indicated. Adapted from Ref. (Merino and Ahedo, 2015) with permission.

As a side note, observe that in MNs with warm or hot ions, the ion thermal energy is also a driver of the expansion. This is relevant, in particular, for thrusters such as the AFMPD and the VASIMR, whose sources generate energetic ions. The parallel ion thermal energy is converted to directed kinetic energy of ions gasdynamically (i.e., just as the thermal energy in a conventional gas is converted to directed energy in an expansion to vacuum). The perpendicular ion thermal energy can be converted to parallel energy by the magnetic mirror force on ions, if ions are sufficiently magnetized (which is the situation expected in the VASIMR operation on hydrogen or other light propellants), or by an electrostatic mirror effect, resulting from a radial potential well around the main plume that forms to keep ions with a large perpendicular inside it (Merino and Ahedo, 2015; Little and Choueiri, 2019). Magnetized ions with initial swirl at the throat can have a diamagnetic azimuthal current that contributes positively as (ion) magnetic thrust.

Far downstream, the plasma must separate from the closed field lines to form a free jet. Otherwise, if the plasma continued attached to the lines, it would return back to the thruster along them, and no momentum would be ejected from the system. Except for huge magnetic strengths and light propellants, ions quickly become effectively unmagnetized downstream. As ions accelerate, their inertial term $m_i u_i^2$ increases, and it was shown that the plasma does not have enough authority to generate the large electric field that would be required to deflect their trajectories to match the magnetic lines (Merino and Ahedo, 2014). As a result, ion trajectories become essentially straight and detach from the applied field. As it can be observed in Fig. 24, any form of electron cooling results in smaller electric fields downstream and an earlier separation of the ion flow. Thus, electron thermodynamics are also central to this key issue.

Electrons, on the contrary, can remain magnetized and follow the magnetic lines far downstream: the difference between the ion and electron velocity directions gives rise to a small but nonzero differential current in the meridional plane, even when the plasma jet carries no net current globally (i.e., there is no local current ambipolarity in the plasma). Electron demagnetization remains an open problem in MN theory. However, it is not expected to affect much thrust generation, since electrons are basically a confined population.

Another phenomenon that gains importance downstream is the influence of the plasma-induced magnetic field (Merino and Ahedo, 2016). This diamagnetic field tends to open the MN lines, increasing the divergence of the jet, and lowers the strength of the net field near the MN axis, facilitating demagnetization. The larger the plasma beta parameter at the throat (i.e. the plasma to magnetic pressure ratio, the earlier in the expansion its effects can be noticed.

## B. Collisionless electron cooling and kinetic effects

The basic plasma/MN model discussed above assumes a cooling of the electron population, according to the polytropic law (Eq. (6)). Experimental data, reviewed in Section 2, suggests fitting it with a polytropic coefficient $\gamma = 1.2 \pm 0.1$. While this electron model is useful to characterize approximately the plasma beam expansion and the total potential fall in the divergent MN, it does not reveal the physics behind the electron cooling.

The polytropic law (Eq. (6)) provides indeed a closure to the fluid equation hierarchy and, in particular, substitutes the use of the electron energy equation (Ahedo *et al*., 2020). Still staying within a conventional fluid formulation, this equation, in the inertialess and stationary case, can be expressed as

$$\nabla \cdot \left[\frac{5}{2}T_e n_e \vec{u}_e + \vec{q}_e\right] \simeq \vec{u}_e \cdot \nabla p_e - Q_{inel}, \tag{11}$$

where $\vec{q}_e$ is the heat flux and $Q_{inel}$ represents inelastic losses due to ionization and excitation of atoms. Neglecting $Q_{inel}$ in the near-collisionless limit, and postulating an adiabatic behavior (*i.e.* $\vec{q}_e = \vec{0}$), Eq. (11) is equivalent to

$$\nabla \ln p_e = \gamma \nabla \ln n_e, \tag{12}$$

with $\gamma = 5/3$. If instead, we postulate a 'convective' behavior of the heat flux, expressed by (Ahedo *et al*., 2020)

$$\vec{q}_e = \alpha \frac{5}{2} T_e n_e \vec{u}_e, \tag{13}$$

with the constant $\alpha$ being the ratio between heat and enthalpy flux, Eq. (12) continues to be fulfilled but with

$$\gamma = \frac{5+5\alpha}{3+5\alpha}. \tag{14}$$

For instance, one has $\gamma = 1.2$ for $\alpha = 1.4$. The convective-type law (Eq. (13)) for the heat flux is far different from the conductive Fourier's law expected in conventional, collisional fluids. However, this type of law has already been suggested in other weakly-collisional plasmas such as divertor plasmas in tokamaks (Stangeby *et al.* ,2010) and laser-produced plasmas (Malone *et al.* ,1975).

Nonetheless, Eq. (13) is just as phenomenological a law as Eq. (6), still not explaining the real physics behind electron cooling. To analyze it requires to acknowledge the near-collisionless character of the electron population, which prevents local thermodynamic equilibrium, thus yielding likely a non-Maxwellian EVDF.

In a weakly-collisional framework, the EVDF satisfies the Boltzmann equation (or the Vlasov equation in the collisionless limit) in the six-dimensional phase space $(\vec{x}, \vec{v})$, with $\vec{v}$ the particle velocity. Macroscopic plasma magnitudes are obtained from integral moments of the EVDF and they satisfy the macroscopic fluid equations, which are also integral moments of the Boltzmann equation. The lack of local thermodynamic equilibrium makes the local solution depend on the global configuration (geometry, magnetic topology, boundary conditions, …) of the problem and is amenable to analytical treatment only in simple configurations.

Martinez-Sanchez *et al*. (2015) studied the kinetic expansion of a collisionless, fully-magnetized plasma along a paraxial (i.e quasi-1D) convergent-divergent MN. A paraxial model solves the distribution functions of ions and electrons along the centerline of the MN, considering the variation of the magnetic field strength, which turns out to be equivalent to the variation of the (inverse of the) area of the magnetic streamtube containing the plasma beam. The stationary model considers the distriubutions of ions and electrons in a far-upstream reservoir, the constants of motion of the distributions emanating from the Vlasov equation (the mechanical energy $E$ and the magnetic moment $\mu$) and the axial electric field to determine the distributions at any spatial location and, then, the macroscopic plasma magnitudes.

The profile of the electrostatic potential along the paraxial MN, $\phi(z)$, is obtained from the quasineutrality condition

$$\int d^3\vec{v} f_i(z,\vec{v}) = \int d^3\vec{v} f_e(z,\vec{v}), \tag{15}$$

which must be solved iteratively, since the ion and electron distributions depend on $\phi(z)$. At each location $z$, it must be determined which ions and electrons, travelling downstream or upstream, can reach that location. For all upstream distributions analyzed so far, $\phi(z)$ is monotonic decreasing from $\phi = 0$ upstream to $\phi = \phi_\infty (< 0)$ downstream, which facilitates that determination. Finally, assuming a current-free plasma beam, the total potential fall $|\phi_\infty|$ in the MN is self-adjusted in the same way that the potential fall is adjusted in a non-neutral Debye sheath in front of a dielectric wall: the value of $|\phi_\infty|$ does not change the net ion current but it controls very effectively the net electron current, so $|\phi_\infty|$ self-adjusts to satisfy the current-free condition. This explains that $|e\phi_\infty|$ depends very much on the properties of the EVDF, and typically it amounts to several times the upstream electron temperature. Sample solutions of the electrostatic potential profile $\phi$ in a current-free MN and two current-carrying MNs are shown in figure 25.

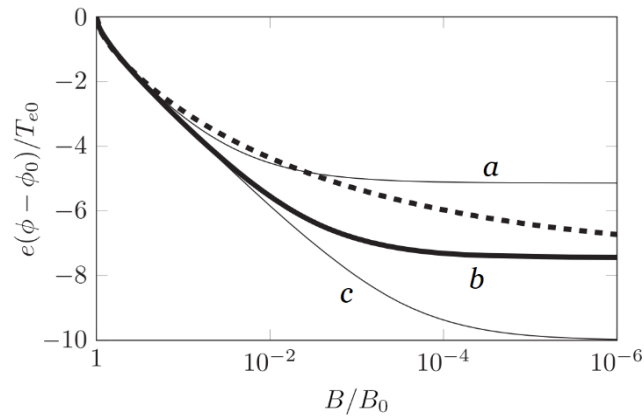

FIG. 25. Dimensionless electrostatic potential $e(\phi - \phi_0)/T_{e0}$ (normalized with $T_{e0}$, the electron temperature at the throat) along the axis of a divergent MN, plotted against the relative magnetic field strength $B/B_0$. Xe as propellant. The solid lines represent kinetic simulations with a net current $j/(en_{e0}\sqrt{T_{e0}/m_i}) = -7$ (a), 0 (b), and $+0.9$ (c). The dotted line represents the polytropic model that results in the same potential fall far downstream as the globally current-free kinetic result (thick line b). Adapted from (Merino *et al*., 2021) with permission.

In the paraxial convergent-divergent MN, the axial motion of an individual ion or electron with energy $E$ and magnetic moment $\mu$ is determined by the electrostatic and the magnetic mirror forces, according to

$$\frac{1}{2}m_\alpha v_z^2 = E - \mu B(z) - Z_\alpha e\phi(z), \tag{16}$$

where $\alpha = i, e$, and $Z_\alpha = \pm 1$ is the charge number of the species. The electrostatic field accelerates ions axially in the convergent and divergent MN sides and decelerates electrons in the two sides. On the contrary, the magnetic mirror decelerates both ions and electrons in the convergent side and accelerates them axially in the divergent side. Combining the electrostatic and magnetic mirror effects, the following situations take place: Upstream ions with high $\mu$ and low $E$ are reflected back to the reservoir within the convergent side, while any ion reaching the MN throat is accelerated downstream (explaining that $|\phi_\infty|$ has no control on the ion current). Therefore, the population of ions in the reservoir is divided into *free* and *reflected* subpopulations, and all ions in the divergent side are free. Regarding electrons from the reservoir, similar subpopulations exist, but only a narrow interval of high $E$ and low $\mu$ constitute the *free* electron subpopulation, even in the divergent side. The main novelty is the existence of a third subpopulation of *doubly-trapped* electrons. These are electrons that bounce back and forth axially between two locations in the divergent MN side (Martinez-Sanchez *et al*., 2015). In their downwards trip, they are accelerated by the magnetic mirror and decelerated by the electric field, and *vice versa* in their upstream trip. Since the trajectories of these electrons are disconnected from the upstream reservoir, their population cannot be determined by the stationary model. Different postulates on this population lead to different expansion gradients and collective electron cooling (Ramos *et al*., 2018). Figure 26 illustrates the EVDF and its different subdomains in the convergent and the divergent side of a MN, when the doubly-trapped electron region in the divergent side is assumed to have the same distribution as the rest of the electrons.

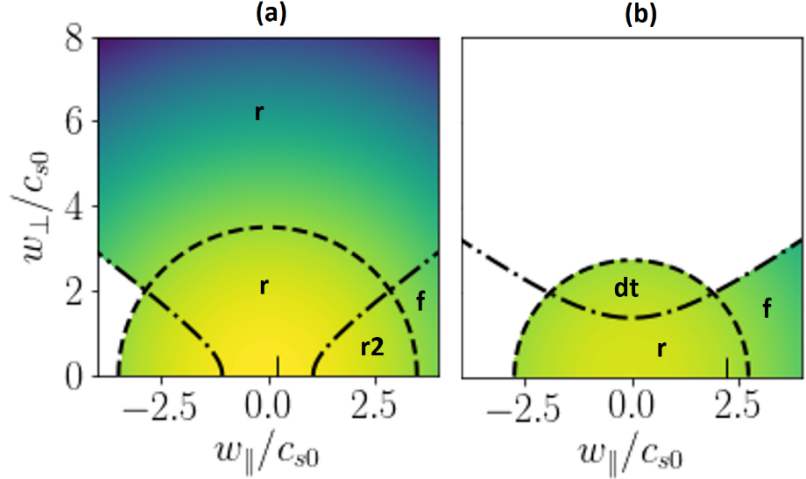

FIG. 26. Electron velocity distribution function (EVDF) in the convergent side (a) and divergent side (b) of a MN. Regions of reflected (r), doubly-trapped (dt) and free (f) electrons are indicated. In the convergent side, reflected electrons that trespass into the divergent side are indicated as (r2). Limiting lines refer to the value of the effective potential on the axial motion of electrons at the throat (dash-dot line) and at infinity downstream (dashed line). Figure adapted from (Ahedo *et al.*, 2020) with permissions; refer to that work for a detailed discussion of this kinetic simulation of the EVDF.

In a collisionless plasma, temperatures are just a measure of the velocity dispersion of each species and generally they are not settled locally. Assuming upstream Maxwellian distributions of ions and electrons, both temperature anisotropy and cooling of ions and electrons develop along the convergent-divergent MN. Temperature anisotropy is related mainly to magnetic mirror effects. These are well known on a single particle, but collective magnetic mirror effects are subtler (Ahedo *et al.*, 2020). For instance, in the MN convergent side, the ratio $T_{\perp i}/T_{\parallel i}$, between ion perpendicular and parallel temperatures, increases much, but $T_{\perp e}/T_{\parallel e}$ remains close to 1 (and thus $f_e$ close to Maxwellian), so the collective magnetic mirror 'force' is strong on ions and near null on electrons at that MN side. On the contrary, in the MN divergent side, the magnetic mirror makes both $T_{\perp i}/T_{\parallel i}$ and $T_{\perp e}/T_{\parallel e}$ to decrease and tend to zero. The behavior of the temperature of each electron subspecies is shown in Fig. 27. These disparities indicate that the combined effects of the magnetic mirror and the electric field redistribute very differently ions and electrons within the EVDF's $\vec{v}$-phase-space. Electron cooling in the MN divergent side is mainly the consequence of the shrinking of the EVDF's $\vec{v}$-phase-space attainable by electrons, as shown in (Martínez-Sánchez *et al.*, 2015; Ahedo *et al.*, 2020) and in Fig. 26.

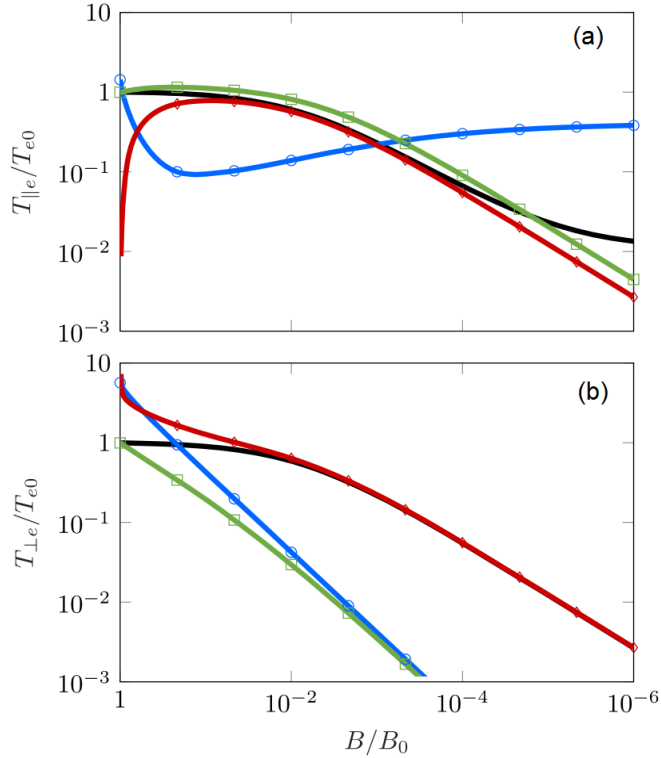

FIG. 27. Parallel (a) and perpendicular (b) temperature of represent free (blue circles), reflected (green squares), doubly-trapped (red diamonds) electrons in a divergent MN, under the assumption that the doubly-trapped regions of the EVDF have the same distribution as the rest of the electrons. Adapted from (Merino *et al.* 2021) with permission.

In a divergent MN, ions are free and constitute a single population that becomes hypersonic downstream, so the particularities of the ion temperatures are not very relevant (there are still some differences in the physical response if upstream ions are hot or cold (Martinez-Sanchez *et al.*, 2015; Ahedo *et al.*, 2020). The situation is very different for electrons, which are constituted by a mixture of the three subpopulations, which have different properties. Doubly-trapped electrons are nearly isotropic, but free and reflected subpopulations are anisotropic; also, free electrons are hotter downstream than the two confined subpopulations. The properties of the resulting electron mixture comes out from weighing the properties of the three subpopulations with their partial densities. This explains that simple physical laws (e.g. in the form of a polytropic equation) are elusive for the electron mixture and highlights the importance of determining correctly the amount of double-trapped electrons and their distribution. The theory of (Martínez-Sánchez *et al.*, 2015) and (Ahedo *et al.*, 2020) postulated that the phase region of double-trapped electrons was fully populated.

There are at least three mechanisms giving access to the doubly-trapped electron region (DTER), enabling filling it (partially) up. One is during the transient formation of the MN, a second one is due to occasional collisional events that bring electron into that region, and a third one is electron-related instabilities. The first one alone leads to a transient-dependent stationary solution, whereas any presence of the latter two would relax the solution slowly toward to a single steady state one. Sanchez-Arriaga *et al.* (2018) developed a time-dependent, direct-Vlasov code of the paraxial MN to assess the transient problem. Contrary to the stationary model relying just on integral equations, the Poisson's equation needs to be solved and the MN domain for numerical integration is finite, which poses some difficulties on the downstream boundary conditions. They find a relatively low fill-up fraction of the DTER compared to the full DTER postulated in (Martinez-Sanchez *et al.*, 2015; Ahedo *et al.*, 2020). This difference has practically no implications on the total potential fall, but it does have them on the expansion plasma profiles and the level of electron cooling (since, as explained above, the electron temperatures are weighted averages over the three subpopulations).

Zhou *et al.* (2021) extended the model (Sanchez-Arriaga *et al*, 2018) to the weak-collisional regime using a BGK approach, which affects almost exclusively to trapped particles. They demonstrate that the fill level of the DTER increases with the effective electron collision frequency but, as long as collisions are scarce compared to the typical electron bouncing time in the DTER, there is not a complete fill, since electrons can flow it and out of that region until an equilibrium is reached. In this weakly-collisional regime, collisions tend to decrease the temperature anisotropy and the electron cooling is more moderate, also more in line with experimental data. Importantly, collisions erase the transient history of the MN formation making the stationary solution unique. It must be noted that reaching a stationary state is rather costly computationally in the weak-collisional regime, as the characteristic time of convergence toward this solution scales inversely with the collision frequency.

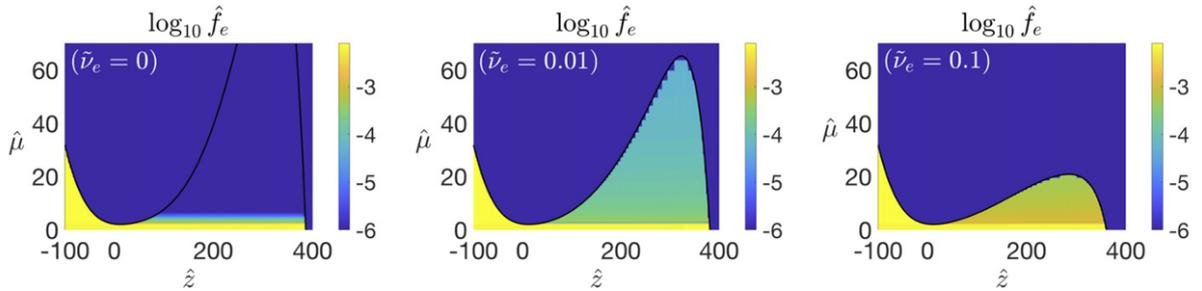

FIG. 28. Dimensionless EVDF $\hat{f}_e(\hat{\mu}, \hat{E})$ in a particular MN as the collision frequency is increased, obtained with the code of (Zhou *et al.*, 2021). The EVDF is presented as a function of the normalized magnetic moment $\hat{\mu} = \mu B_0/T_{e0}$ and energy $\hat{E} = E/T_{e0}$. These plots correspond with energy $\hat{E} = 2.85$. The black line delimits the allowed region from the forbidden region in this phase space. Adapted from (Zhou *et al.*, 2021) with permission.

Once the basic physics of electron temperature anisotropy and cooling have been established, the question of whether a reliable macroscopic electron (and ion) model for the MN weak-collisionality scenario can be derived, usable within the fluid formalism, remains open. For the paraxial MN model and no collisions between two particles of different species, the kinetic solution for species $j = i, e$, satisfies the following macroscopic equations,

$$\frac{n_j u_j}{B} = const, \tag{17}$$

$$m_j n_j u_j \frac{du_j}{dz} + Z_j e \frac{d\phi}{dz} + \frac{d(n_j T_{\parallel j})}{dz} + n_j(T_{\perp j} - T_{\parallel j})\frac{d\ln\ln B}{dz} = 0, \tag{18}$$

$$\left(Z_j e\phi + \frac{1}{2}m_j u_j^2 + h_j\right)\frac{n_j u_j}{B} + \frac{q_j}{B} = const, \tag{19}$$

$$\frac{T_{\perp j} n_j u_j + q_{\perp j}}{B^2} = C_j \tag{20}$$

Here, $u_j$ is the macroscopic velocity parallel to the magnetic line, $Z_j$ is the species charge number, $h_j = \frac{3T_{\parallel j}}{2} + T_{\perp j}$ is the enthalpy per particle, $q_j = \frac{m_j}{2}\int d^3\vec{v} f_e c_{jz} c_j^2$, $\vec{c}_j = \vec{v} - u_j \vec{1}_z$, is the heat flux parallel to the magnetic line, $q_{\perp j} = \frac{m_j}{2}\int d^3\vec{v} f_e c_{jz} c_{\perp j}^2$ is the heat flux, parallel to the magnetic line, of perpendicular energy, and $C_j$ a function dependent on intraspecies collisions (it is constant in the collisionless limit).

Equation (17) expresses the conservation of the species flow, with $1/B$ being proportional to the area of the plasma beam. The momentum equation (Eq. (18)) – along the magnetic line – illustrates that a collective magnetic mirror effect is intimately linked to the development of temperature anisotropy. Equation (19) expresses the conservation of total energy, with $Z_j e\phi$ and $h_j + \frac{q_j}{n_j u_j}$ the potential and thermal energy, per particle, respectively. Equation (20) sets the conservation of perpendicular energy. Its simple form is because it refers only to the MN centerline where there are no perpendicular gradients. The dependence on $1/B^2$ explains that the perpendicular thermal energy flux $T_{\perp j} n_j u_j + q_{\perp j}$ goes to zero at $B \to 0$.

The closure of the set Eqs. (17) – (20) requires defining laws for the heat fluxes $q_j$ and $q_{\perp j}$, in terms of low-order magnitudes ($n_j, u_j, T_{\parallel j}, T_{\perp j}$) and independent of fourth-order integral moments. For a collisional species, $q_{\perp j} = 2q_j/3$, and $q_j$ satisfies the conductive Fourier law. For weakly-collisional species, simple laws are not going to exist in general, so the best that can be expected are approximate laws for particular regions of the discharge. Centering the attention on the electrons and the expansion in the divergent MN, (Ahedo *et al.*, 2020) and (Zhou *et al.*, 2021) showed that $q_{\perp e}$ can be neglected, and $q_e$ does not follow a conductive (i.e. Fourier) law $q_e \propto dT_e/dz$. Instead, the convective-type law (Eq. (13)) offers an acceptable approximation.

In the context of tokamaks and plasma-laser applications and in order to cover intermediate collisional regimes, (Stangeby *et al.*, 2010), (Bell *et al.*, 1985), and (Zawaideh *et al.*,1988) proposed hybrid closure laws for the heat flux with one fitting parameter. Following this approach Zhou *et al*. (2021) has attempted to fit the heat flux in the divergent MN nozzle with the two-parameter hybrid law

$$\vec{q}_e = \underline{\alpha} T_e n_e \vec{u}_e - \underline{\beta} K \nabla T_e, \tag{21}$$

with $K$ the thermal conductivity, and $\underline{\alpha}$ and $\underline{\beta}$ two fitting parameters, that depend on the electron collisionality. The kinetic results demonstrated that, in the evolution from a collisionless scenario to a collisional one, $\alpha$ goes from $O(1)$ to 0 and $\beta$ from 0 to 1, which supports the reliability of this hybrid approach.

The kinetic studies commented on so far are limited to the paraxial, fully-magnetized MN model. Merino *et al*. (2021) have recently extended the fully-magnetized case to the 2D configuration, where still the response in each magnetic line can be tackled independently. Anisotropy, cooling, and parallel heat flux follow exactly the same trends while plasma parameters adapt to radially varying boundary conditions.

Partial magnetization of ions leads to a more complex 2D problem, still unsolved with kinetic electrons, but no fundamental changes are expected in the electron response. However, freeing, far downstream, the full-magnetization postulate on electrons, which certainly happens in any real MN, can imply important changes. The effect of this gradual demagnetization on $T_{\parallel e}$ and $T_{\perp e}$ is unknown. Nonetheless, studies by Merino *et al*. (2018) on an unmagnetized, collisionless, paraxial, plasma plume, with electrons under electrostatic confinement only, show a collective behavior very similar to the one in the MN in terms of anisotropy and cooling of the electron temperature, at least around the plume axis. Instead of performing gyro orbits around magnetic lines, electrons perform large excursions in a radial electrostatic potential well, and this brings about an 'electrostatic mirror' effect that plays an analogous similar role to the magnetic mirror effect.

## IV. CONCLUSIONS AND FUTURE CHALLENGES

In this review, we discussed the fundamental physics of the kinetics of electron cooling in MNs and magnetically expanding plasmas. When considering the actual plasma devices using the power source, a non-local coupling of the RF or MW power with electrons often occur, e.g., via a wave-heating mode; the adiabatic conditions cannot be maintained. The investigation of the polytropic index under the experiments having a heat source and loss would also contribute to understand the energy transfer process by combining with the detailed physical process; thus, well controlled experiments will be required to investigate such a process in laboratories.

On the theoretical side, it has been argued that, while fluid models enable a sufficient description of certain MN physical mechanisms, there are currently no self-consistent closure laws for the non-local electron thermodynamics in quasi-collisionless regimes. While physically unjustified, the closure most commonly used is an isotropic polytropic law, with an exponent $\gamma$ that is used to fit the observed electron cooling in experimental results. As the electric fields that accelerate ions in the MN are proportional to the local value of the electron temperature, obtaining a correct description of electron cooling and anisotropy is essential for the correct prediction of MN performance figures and ion detachment downstream.

A paraxial kinetic model of electrons has been reviewed, which enables the self-consistent solution of the EVDF and electrostatic potential, and therefore the self-consistent solution of electron cooling and anisotropy development. Electrons are seen to divide into free, reflected, and doubly-trapped electrons depending on their location along the MN, their energy, and their magnetic moment. Doubly-trapped electrons cannot be from a the steady-state, collisionless description of the problem, and require tackling the transient and/or including the effect of the small collisionality in the MN. The total potential fall along the plume is intimately linked to the amount of free electrons and the net electron current in the device. Each electron subpopulation cools down differently along the expansion and an initially isotropic EVDF becomes anisotropic downstream. While one can define a polytropic model that results in the same total potential fall as the globally current-free kinetic model, the map of the electrostatic potential differs substantially, and the anisotropy is missed.

More advances closure laws may need to resort to modeling the electron heat fluxes in a way that they respect the kinetic solution of the plasma expansion. Other open challenges in the modeling field include the self-consistent simulation of 2D MNs and the study of electron demagnetization and detachment. Given the large computational cost of a direct kinetic simulation, smart approaches such as hybrid combinations of fluid and kinetic descriptions to lower the number of numerical operations may offer a way forward in this area.

**ACKNOWLEDGMENTS**

June Young Kim's work was supported by the Basic Science Research Program through the National Research Foundation of Korea (NRF) funded by the Ministry of Science, ICT, and Future Planning (Grant No. 2020R1C1C1009547) and partly by the NRF funded by the Korean government (MSIT) (Grant No. 2019M2D1A1080261).

Kazunori Takahashi's work was supported by the Grant-in- Aid for Scientific Research (Grant No. 19H00663 and 21K18611) from the Japan Society for the Promotion of Science, Fusion Oriented REsearch for disruptive Science and Technology (FOREST) from Japan Science and Technology Agency (Grant No. JPMJFR212A).

Mario Merino's work was supported by the European Research Council (ERC) under the European Union's Horizon 2020 research and innovation programme (project ERC Starting Grant ZARATHUSTRA, grant agreement No 950466).

Eduardo Ahedo's work is part of project I+D+i PID2019-108034RB-I00, funded by MCIN/ AEI/10.13039/501100011033/.

**AUTHOR DECLARATIONS**

**Conflict of Interest**

The authors have no conflicts to disclose.

**Author Contributions**

June Young Kim: Supervision (lead); Formal analysis (lead); Investigation (lead); Writing – original draft (lead); Writing – review & editing (lead). Kyoung-Jae Chung: Formal analysis (supporting); Writing – review & editing (supporting). Kazunori Takahashi: Validation (supporting); Writing – original (supporting); Writing – review & editing (supporting). Mario Merino: Investigation (equal); Writing – original draft (equal); Writing – review & editing (equal). Eduardo Ahedo: Investigation (supporting); Writing – original draft (equal); Writing – review & editing (supporting).

**DATA AVAILABILITY**

The data that support the findings of this study are available from the corresponding authors upon reasonable request

**REFERENCES**

Ahedo, E. and Martinez-Sanchez M., 'Theory of a Stationary Current-Free Double Layer in a Collisionless Plasma', Physical Review Letters 103, 135002 (2009)

Ahedo, E. and Merino, M., "Two-dimensional supersonic plasma acceleration in a magnetic nozzle," Physics of Plasmas 17, 073501 (2010).

Ahedo, E., "Plasmas for space propulsion," Plasma Physics and Controlled Fusion 53, 124037 (2011a).

Ahedo E., 'Double-layer formation and propulsive assessment for a three-species plasma expanding in a magnetic nozzle', Physics of Plasmas 18, 033510 (2011b)

Ahedo, E. and Merino, M., "On plasma detachment in propulsive magnetic nozzles," Physics of Plasmas 18, 053504 (2011).

Ahedo, E. and Merino, M., "Two-dimensional plasma expansion in a magnetic nozzle: Separation due to. electron inertia," Physics of Plasmas 19, 083501 (2012).

Ahedo, E., Correyero, S., Navarro-Cavallé, J. and Merino, M., "Macroscopic and parametric study of a kinetic. plasma expansion in a paraxial magnetic nozzle," Plasma Sources Science and Technology 29, 045017 (2020).

Andersen, S.A., Jensen, V.O., Nielsen, P. and D'Angelo, N., "Continuous supersonic plasma wind tunnel," The. Physics of Fluids 12, 557-560 (1969).

Andrenucci, M., "Magnetoplasmadynamic thrusters," Encyclopedia of Aerospace Engineering (2010).

Arefiev, A.V. and Breizman, B.N., "Theoretical components of the VASIMR plasma propulsion concept," Physics of Plasmas 11, 2942-2949 (2004).

Arefiev, A.V. and Breizman, B.N., "Magnetohydrodynamic scenario of plasma detachment in a magnetic nozzle," Physics of Plasmas 12,043504 (2005).

Arefiev, A.V. and Breizman, B.N., "Ambipolar acceleration of ions in a magnetic nozzle," Physics of Plasmas. 15, 042109 (2008).

Bell, A.R., 'Non-Spitzer heat flow in a steadily ablating laser-produced plasma', Physics of Fluids 28, 2007 (1985)

Boni, F., Jarrige, J. and Désangles, V., "Plasma expansion and electron properties in a magnetic nozzle. electrode-less thruster,". International Electric Propulsion Conference 2022

Boswell, R.W., Takahashi, K., Charles, C. and Kaganovich, I.D., "Non-local electron energy probability. function in a plasma expanding along a magnetic nozzle," Frontiers in Physics 3, 14 (2015).

Breizman, B.N., Tushentsov, M.R. and Arefiev, A.V., "Magnetic nozzle and plasma detachment model for a steady-state flow," Physics of Plasmas 15, 057103 (2008).

Charles, C., Boswell, R.W., Cox, W., Laine, R. and MacLellan, P., "Magnetic steering of a helicon double layer. thruster," Applied Physics Letters 93, 201501 (2008).

Charles, C., "High density conics in a magnetically expanding helicon plasma," Applied Physics Letters 96, 051502 (2010).

Chang-Diaz, F., Squire, J., Bengtson, R., Breizman, B., Carter, M. and Baity, F., "The physics and engineering. of the VASIMR engine," 36th AIAA/ASME/SAE/ASEE Joint Propulsion Conference and Exhibit, 3756 (2000).

Chang-Diaz, F., "The VASIMR rocket. Scientific American," 283, 90-97 (2000).

Choueiri, E., "Scaling of thrust in self-field magnetoplasmadynamic thrusters," Journal of Propulsion and Power 14, 744-753 (1998).

Correyero, S., Jarrige, J., Packan, D. and Ahedo, E., "Plasma beam characterization along the magnetic nozzle. of an ECR thruster, Plasma Sources Science and Technology 28, 095004 (2019).

Deline, C.A., Bengtson, R.D., Breizman, B.N., Tushentsov, M.R., Jones, J.E., Chavers, D.G., Dobson, C.C. and. Schuettpelz, B.M., "Plume detachment from a magnetic nozzle," Physics of Plasmas 16, 033502 (2009).

Ebersohn, F., Girimaji, S., Staack, D., Shebalin, J., Longmier, B. and Olsen, C., "Magnetic nozzle plasma. plume: review of crucial physical phenomena," 48th AIAA/ASME/SAE/ASEE Joint Propulsion Conference & Exhibit, 4274 (2012).

Ghosh, S., Yadav, S., Barada, K.K., Chattopadhyay, P.K., Ghosh, J., Pal, R. and Bora, D., "Formation of annular plasma downstream by magnetic aperture in the helicon experimental device," Physics of Plasmas 24, 020703 (2017).

Gulbrandsen, N. and Fredriksen, Å., "RFEA measurements of high-energy electrons in a helicon plasma device. with expanding magnetic field," Frontiers in Physics 5, 2 (2017).

Hepner, S., Wachs, B. and Jorns, B., "Wave-driven non-classical electron transport in a low temperature. magnetically expanding plasma," Applied Physics Letters 116, 263502 (2020).

Hepner, S.T. and Jorns, B., "Anomalous electron thermal conductivity in a magnetic nozzle," AIAA Propulsion. and Energy 2021 Forum, 3399 (2021).

Hooper, E.B., "Plasma detachment from a magnetic nozzle," Journal of Propulsion and Power 9, 757-763. (1993).

Hu, Y., Huang, Z., Cao, Y. and Sun, Q., "Kinetic insights into thrust generation and electron transport in a. magnetic nozzle," Plasma Sources Science and Technology 30, 075006 (2021).

Imai, R. and Takahashi, K., "Demonstrating a magnetic steering of the thrust imparted by the magnetic nozzle. radiofrequency plasma thruster" Applied Physics Letters 118, 264102 (2021).

Kaganovich, I.D., Smolyakov, A., Raitses, Y., Ahedo, E., Mikellides, I.G., Jorns, B., Taccogna, F., Gueroult, R., Tsikata, S., Bourdon, A. and Boeuf, J.P., "Physics of E× B discharges relevant to plasma propulsion and similar technologies," Physics of Plasmas 27, 120601 (2020).

Kim, J.Y., Chung, K.S., Kim, S., Ryu, J.H., Chung, K.J. and Hwang, Y.S., "Thermodynamics of a magnetically. expanding plasma with isothermally behaving confined electrons," New Journal of Physics 20, 063033 (2018).

Kim, J.Y., Jang, J.Y., Chung, K.S., Chung, K.J. and Hwang, Y.S., "Time-dependent kinetic analysis of trapped. electrons in a magnetically expanding plasma," Plasma Sources Science and Technology 28, 07LT01 (2019).

Kim, J.Y., Go, G., Hwang, Y.S. and Chung, K.J., "Dependence of the polytropic index of plasma on magnetic. field" *New Journal of Physics 23*, 052001 (2021a).

Kim, J.Y., Lee, H.C., Go, G., Choi, Y.H., Hwang, Y.S. and Chung, K.J., "Exploring the nonextensive. thermodynamics of partially ionized gas in magnetic field," Physical Review E 104, 045202 (2021b).

Kim, J.Y., Jang, J.Y., Choi, J., Wang, J.I., Jeong, W.I., Elgarhy, M.A.I., Go, G., Chung, K.J. and Hwang, Y.S., "Magnetic confinement and instability in partially magnetized plasma," Plasma Sources Science and Technology 30, 025011 (2021c).

Kuriki, K. and Okada, O., "Experimental study of a plasma flow in a magnetic nozzle," The Physics of Fluids 13, 2262-2269 (1970).

Kodys, A. and Choueiri, E., "A critical review of the state-of-the-art in the performance of applied-field magnetoplasmadynamic thrusters," 41st AIAA/ASME/SAE/ASEE Joint Propulsion Conference & Exhibit, 4247 (2005).

Lafleur, T., Cannat, F., Jarrige, J., Elias, P.Q. and Packan, D., "Electron dynamics and ion acceleration in. expanding-plasma thrusters," Plasma Sources Science and Technology 24, 065013 (2015).

Levchenko, I., Xu, S., Mazouffre, S., Lev, D., Pedrini, D., Goebel, D., Garrigues, L., Taccogna, F. and Bazaka, K., "Perspectives, frontiers, and new horizons for plasma-based space electric propulsion," Physics of Plasmas 27, 020601 (2020).

Little, J.M. and Choueiri, E.Y., "Electron cooling in a magnetically expanding plasma," Physical Review. Letters 117, 225003 (2016).

Little, J.M. and Choueiri, E.Y., "Electron demagnetization in a magnetically expanding plasma," Physical. Review Letters 123, 145001 (2019).

Little, J.M. and Choueiri, E.Y., "Thrust and efficiency model for electron-driven magnetic nozzle," Physics of. Plasmas 20,103501 (2013).

Litvinov, I.I., "Stationary efflux into a vacuum by a dual-temperature fully ionized plasma," Journal of Applied. Mechanics and Technical Physics 12, 793-802 (1971).

Longmier, B.W., Bering, E.A., Carter, M.D., Cassady, L.D., Chancery, W.J., Díaz, F.R.C., Glover, T.W., Hershkowitz, N., Ilin, A.V., McCaskill, G.E. and Olsen, C.S., "Ambipolar ion acceleration in an expanding magnetic nozzle," Plasma Sources Science and Technology 20, 015007 (2011).

Malone, R. C., McCrory, R. L., and Morse, R. L., 'Indications of strongly flux-limited electron thermal conduction in laser-target experiments', Physical Review Letters 34, (1975)

Martinez-Sanchez, M., Navarro-Cavallé, J. and Ahedo, E., "Electron cooling and finite potential drop in a. magnetized plasma expansion," Physics of Plasmas 22, 053501 (2015).

Merino, M. and Ahedo, E., "Plasma detachment in a propulsive magnetic nozzle via ion. demagnetization," Plasma Sources Science and Technology 23, 032001 (2014).

Merino, M. and Ahedo, E., 2016. Magnetic nozzles for space plasma thrusters.

Merino, M. and Ahedo, E., "Space plasma thrusters: Magnetic nozzles for," Encyclopedia of Plasma Technology 2, 1329-1351 (2017).

Merino, M., Mauriño, J. and Ahedo, E., "Kinetic electron model for plasma thruster plumes," Plasma Sources. Science and Technology 27, 035013 (2018).

Merino, M., Nuez, J. and Ahedo, E., "Fluid-kinetic model of a propulsive magnetic nozzle," Plasma Sources. Science and Technology 30, 115006 (2021).

Merino, M. and Ahedo, E., "Two-dimensional quasi-double-layers in two-electron-temperature, current-free. plasmas," Physics of Plasmas 20, 023502 (2013).

Merino, M. and Ahedo, E., "Influence of electron and ion thermodynamics on the magnetic nozzle plasma. expansion," IEEE Transactions on Plasma Science 43, 244-251 (2015).

Merino, M. and Ahedo, E., "Effect of the plasma-induced magnetic field on a magnetic nozzle," Plasma Sources. Science and Technology 25, 045012 (2016).

Merino, M. and Ahedo, E., "Contactless steering of a plasma jet with a 3D magnetic nozzle," Plasma Sources Science and Technology 26, 095001 (2017).

Merino, M., Nuez, J. and Ahedo, E., "Fluid-kinetic model of a propulsive magnetic nozzle," Plasma Sources. Science and Technology, 30, 115006 (2021).

Olsen, C.S., Ballenger, M.G., Carter, M.D., Diaz, F.R.C., Giambusso, M., Glover, T.W., Ilin, A.V., Squire, J.P., Longmier, B.W., Bering, E.A. and Cloutier, P.A., "Investigation of plasma detachment from a magnetic nozzle in the plume of the VX-200 magnetoplasma thruster," IEEE Transactions on Plasma Science 43, 252-268 (2014).

Raadu, M.A., "Expansion of a plasma injected from an electrodeless gun along a magnetic field," Plasma. Physics 21, 331 (1979).

Ramos, J.J., Merino, M. and Ahedo, E., "Three dimensional fluid-kinetic model of a magnetically guided. plasma jet," Physics of Plasmas 25, 061206 (2018).

Sánchez-Arriaga, G., Zhou, J., Ahedo, E., Martínez-Sánchez, M. and Ramos, J.J., "Kinetic features and non-stationary electron trapping in paraxial magnetic nozzles," Plasma Sources Science and Technology 27, 035002 (2018).

Sercel, J., "Electron-cyclotron-resonance (ECR) plasma acceleration," 19th AIAA, Fluid Dynamics, Plasma. Dynamics, and Lasers Conference 1407 (1987).

Sheehan, J.P., Longmier, B.W., Bering, E.A., Olsen, C.S., Squire, J.P., Ballenger, M.G., Carter, M.D., Cassady, L.D., Díaz, F.C., Glover, T.W. and Ilin, A.V., "Temperature gradients due to adiabatic plasma expansion in a magnetic nozzle," Plasma Sources Science and Technology 23, 045014 (2014).

Sheppard, A.J. and Little, J.M., "Scaling laws for electrodeless plasma propulsion with water vapor. propellant," Plasma Sources Science and Technology 29, 045007 (2020).

Singh, N., Rao, S. and Ranganath, P., "Waves generated in the plasma plume of helicon magnetic. nozzle," Physics of Plasmas 20, 032111 (2013).

Stangeby, P.C., Canik, J.M., and Whyte, D.G.,'The relation between upstream density and temperature widths in the scrape-off layer and the power width in an attached divertor', Nuclear Fusion 50, 125003 (2010)

Sutton, G.P. and Biblarz, O., "Rocket propulsion elements," John Wiley & Sons (2016).

Takahashi, K., Charles, C., Boswell, R., Cox, W. and Hatakeyama, R., "Transport of energetic electrons in a magnetically expanding helicon double layer plasma," Applied Physics Letters 94, 191503, (2009).

Takahashi, K., Charles, C., Boswell, R.W. and Fujiwara, T., "Electron energy distribution of a current-free. double layer: Druyvesteyn theory and experiments," Physical Review Letters 107, 035002 (2011a).

Takahashi, K., Lafleur, T., Charles, C., Alexander, P., Boswell, R.W., Perren, M., Laine, R., Pottinger, S., Lappas, V., Harle, T. and Lamprou, D., "Direct thrust measurement of a permanent magnet helicon double layer thruster," Applied Physics Letters 98, 141503 (2011b).

Takahashi, K., Akahoshi, H., Charles, C., Boswell, R.W. and Ando, A., "High temperature electrons exhausted. from rf plasma sources along a magnetic nozzle," Physics of Plasmas 24, 084503 (2017a).

Takahashi, K. and Ando, A., "Laboratory observation of a plasma-flow-state transition from diverging to. stretching a magnetic nozzle," Physical Review Letters 118, 225002 (2017b).

Takahashi, K., Charles, C., Boswell, R. and Ando, A., "Adiabatic expansion of electron gas in a magnetic. nozzle," Physical Review Letters 120, 045001 (2018).

Takahashi, K., "Helicon-type radiofrequency plasma thrusters and magnetic plasma nozzles,". Reviews of. Modern Plasma Physics 31, 1-61 (2019).

Takahashi, K., Charles, C., Boswell, R.W. and Ando, A., "Thermodynamic analogy for electrons interacting. with a magnetic nozzle," Physical Review Letters 125, 165001 (2020a).

Takahashi, K., Takao, Y. and Ando, A., "Increased thrust-to-power ratio of a stepped-diameter helicon plasma. thruster with krypton propellant," Journal of Propulsion and Power 36, 961-965 (2020b).

Takahashi, K. and Imai, R., "Two-dimensional deflection of a plasma plume exhausted from a magnetically. steered radiofrequency plasma thruster," Physics of Plasmas 29, 054501 (2022).

Plihon, N., Chabert, P. and Corr, C.S., "Experimental investigation of double layers in expanding. plasmas," Physics of Plasmas 14, 013506 (2007).

Vialis, T., Jarrige, J., Aanesland, A. and Packan, D., "Direct thrust measurement of an electron cyclotron. resonance plasma thruster" Journal of Propulsion and Power 34, 1323-1333 (2018).

Vinci, A.E., Delavière–Delion, Q. and Mazouffre, S., "Electron thermodynamics along magnetic nozzle lines in. a helicon plasma," Journal of Electric Propulsion 1, 1-12 (2022).

Zawaideh, E. and Kim, N.S. and Najmabadi, F., 'Generalized parallel heat transport equations in collisional to weakly collisional plasmas', Physics of Fluids 31, 3280 (1988)

Zhang, Y., Charles, C. and Boswell, R., "Thermodynamic study on plasma expansion along a divergent. magnetic field," Physical Review Letters 116, 025001 (2016a).

Zhang, Y., Charles, C. and Boswell, R., "A polytropic model for space and laboratory plasmas described by bi-Maxwellian electron distributions," The Astrophysical Journal 829, 10 (2016b).

Ziemba, T., Carscadden, J., Slough, J., Prager, J. and Winglee, R., "High power helicon thruster," 41st. AIAA/ASME/SAE/ASEE Joint Propulsion Conference & Exhibit, 4119 (2005).

Zhou, J., Sánchez-Arriaga, G. and Ahedo, E., "Time-dependent expansion of a weakly-collisional plasma beam in a paraxial magnetic nozzle," Plasma Sources Science and Technology 30, 045009 (2021).